\documentclass[aps,11pt,showpacs,showkeys,nofootinbib]{revtex4}
\usepackage{amsmath,amssymb,amsfonts,pifont,fancybox,float}

\usepackage{graphicx}
\usepackage{epstopdf}

\begin{document}
\title{Competing Orders in
One-Dimensional Half-Integer Fermionic Cold Atoms: 
A Conformal Field Theory Approach}
\author{P. Lecheminant}
\email{Philippe.Lecheminant@ptm.u-cergy.fr}
\affiliation{
Laboratoire de Physique Th\'eorique et
Mod\'elisation, CNRS UMR 8089,
Universit\'e de Cergy-Pontoise, Site de Saint-Martin,
2 avenue Adolphe Chauvin,
95302 Cergy-Pontoise Cedex, France}
\author{P. Azaria}
\affiliation{
Laboratoire de Physique Th\'eorique de la Mati\`ere Condens\'ee,
Universit\'e Pierre et Marie Curie, 4 Place Jussieu,
75256 Paris Cedex 05, France}
\author{E. Boulat}
\affiliation{Laboratoire Mat\'eriaux et Ph\'enom\`enes Quantiques, CNRS UMR 7162, Universit\'e Paris Diderot,
75205 Paris Cedex 13, France}
\begin{abstract}
The physical properties of arbitrary half-integer spins $F = N - 1/2$
fermionic cold atoms loaded into a one-dimensional optical lattice are
investigated by means of a conformal field theory approach. 
We show that for attractive interactions 
two different superfluid phases emerge for $F \ge 3/2$:
A BCS pairing phase, and a molecular superfluid phase which 
is formed from bound-states made of $2N$ fermions.
In the low-energy approach, the competition between 
these instabilities and charge-density waves 
is described in terms of ${\mathbb{Z}}_N$ parafermionic degrees of freedom. 
The quantum phase transition for $F=3/2,5/2$ is universal 
and shown to belong to the
Ising and three-state Potts universality classes respectively.  
In contrast, for $F \ge 7/2$, the transition is non-universal.
For a filling of one atom per site, a Mott transition occurs and
the nature of the possible Mott-insulating phases are determined.
\end{abstract}
\maketitle

\section{Introduction}

Ultracold atomic physics has attracted a lot of interest in recent years 
with the opportunity to study strongly correlated effects,
such as high-temperature superconductivity, in a new context \cite{hofstetter}.
In particular, loading cold atomic gases into an optical lattice allows for the
realization of bosonic and fermionic lattice models and the
experimental study of exotic quantum phases~\cite{reviewcold}. 
A prominent example is the toolbox to 
engineer the three-dimensional Bose-Hubbard model \cite{jaksch}
and the observation of its Mott insulator -- superfluid quantum phase transition
with cold bosonic atoms in an optical lattice \cite{greiner}.

Ultracold
atomic systems also  offer an opportunity to investigate the effect of
spin degeneracy since the atomic total angular momentum $F$, which includes
both electron and nuclear spins, can be
larger than $1/2$ resulting in $2F+1$ hyperfine states.  
In magnetic traps, these $2F+1$ components are split, while in optical 
traps these (hyperfine) spin degrees of freedom are degenerate and 
novel interesting phases might emerge.
For instance, Bose-Einstein condensates of bosonic atoms with a nonzero total 
spin $F$ are expected to display rich interesting structures in spin space
like in superfluid $^{3}$He.
In this respect, various exotic superfluid 
condensates (including nematic ones), Mott insulating phases,
and non-trivial vortex structures have 
been predicted recently in spinor bosonic
atoms with $F \ge 1$ \cite{ho,ohmi,zhou,demler,moore,yip,bosonspin3}.  
These theoretical predictions
might be checked in the context of Bose-Einstein condensates of
sodium, rubidium atoms \cite{bosonspin1} 
and in spin-3 atom of $^{52}$Cr~\cite{spin3}.
The spin-degeneracy in fermionic atoms, 
like $^{6}$Li, $^{40}$K or $^{173}$Yb, is also expected to give rise
to some interesting 
superfluid phases \cite{hoyip,honerkamp,paananen,cherng,he}.  
In particular, a
molecular superfluid (MS) phase might be stabilized in multicomponent 
attractive Fermi gas where more than two
fermions form a bound state.
Such a non-trivial superfluid behavior
has already been found in different
contexts.
In nuclear physics, a four-particle condensate like the $\alpha$ particle
is favored over deuteron condensation at low density \cite{schuck} and
it may have implications for light nuclei and asymmetric matter
in nuclear stars \cite{tohsaki,deanlee}.
This quartet condensation can occur in the field of semiconductors
with the formation of biexciton \cite{nozieres}.
A quartetting phase, which stems from the pairing of Cooper pairs, also
appears in a model
of one-dimensional Josephson junctions \cite{doucot}.
A possible experimental observation of quartets 
might be found in superconducting 
quantum interference devices with (100)/(110) interfaces 
of two d-wave superconductors \cite{schneider}. 
In particular, the $hc/4e$ periodicity of 
the critical current with applied magnetic flux 
has been interpreted as the formation of 
quartets with total charge $4e$ \cite{aligia}.
Finally, a bipairing mechanism has also been  predicted in 
four-leg Hubbard ladders \cite{affleckbipairing}.

Recently, the emergence of quartet and 
triplet (three-body bound states) has been proposed to occur in
the context of ultracold 
fermionic atoms \cite{miyake,Wu2005,phle,capponiMS,Rapp,schlottmann}.
Much of these studies have been restricted to the special 
case of an SU($n$) symmetry between the hyperfine states
of a $n$-component Fermi gas.  
In this paper, we investigate the generic physical features of
half-integer spins $F= N - 1/2$ fermionic cold 
atoms with s-wave scattering interactions loaded into
a one-dimensional optical lattice.  The low-energy physical
properties of $2F+1 = 2N$-component fermions with contact interactions
are known to be described by a
Hubbard-like Hamiltonian \cite{ho}:
\begin{equation}
{\cal H} = -t \sum_{i,\alpha} \left[c^{\dagger}_{\alpha,i} c_{\alpha,
i+1} + {\rm H.c.} \right] 
- \mu \sum_i n_i
+ \sum_{i,J} U_J \sum_{M=-J}^{J}
P_{JM,i}^{\dagger} P_{JM,i},
\label{hubbardSgen}
\end{equation}
where $c^{\dagger}_{\alpha,i}$ ($\alpha = 1,..., 2N$) is the fermion
creation operator corresponding to the $2F+1=2N$ atomic states
and $n_i = \sum_{\alpha}  c^{\dagger}_{\alpha,i} c_{\alpha,i}$ is the
density operator on site $i$.  The
pairing operators in Eq. (\ref{hubbardSgen}) are defined through the
Clebsch-Gordan coefficients
 for forming a total 
spin $J$ from two spin-$F$ fermions:
$P^{\dagger}_{JM,i} = \sum_{\alpha \beta} \langle JM|F,F;\alpha
\beta\rangle c^{\dagger}_{\alpha,i}c^{\dagger}_{\beta,i}$.  The
interactions are SU(2) spin-conserving and depend on $U_J$ parameters
corresponding to the total spin $J$ which takes
only even integers value due to Pauli's principle: $J=0,2,...,2N-2$.
Even in this simple scheme the interaction pattern is still involved since
there are $N$ coupling constants in the general spin-$F$ case.
In this paper, we shall consider a two coupling-constant
version of model (\ref{hubbardSgen}) with 
$U_2 = ... = U_{2N-2} \ne U_0$ which 
incorporates the relevant physics of higher-spin
degeneracy with respect to the formation of an exotic MS phase.
To this end, it is enlightening to express 
this model in terms of the density $n_i$
and the BCS singlet-pairing operator for spin-$F$ which is defined by:
\begin{equation}
P^{\dagger}_{00,i} =  \frac{1}{\sqrt{2N}} \sum_{\alpha, \beta}
c^{\dagger}_{\alpha,i} {\cal J}_{\alpha \beta} c^{\dagger}_{\beta,i}
= \frac{1}{\sqrt{2N}} \sum_{\alpha} \left(-1\right)^{2F+\alpha}
c^{\dagger}_{\alpha,i} c^{\dagger}_{2N+1-\alpha,i} ,
\label{bcs}
\end{equation}
where the matrix ${\cal J}$ is the natural generalization of the familiar
antisymmetric tensor $\epsilon = i \sigma_2$ to spin $F > 1/2$.  Such
a singlet-pairing operator has been extensively studied in the context
of two-dimensional frustrated spin-1/2 quantum magnets \cite{sachdev}.
Using the relation  
$\sum_{J,M} P^{\dagger}_{JM,i} P_{JM,i} = n_i^2$,
model (\ref{hubbardSgen}) with
$U_2 = ... = U_{2N-2} \ne U_0$ reads then as follows:
\begin{equation}
{\cal H} = -t \sum_{i,\alpha} [c^{\dagger}_{\alpha,i} c_{\alpha,i+1} +
{\rm H.c.} ]
- \tilde\mu \sum_i n_i
+ \frac{U}{2} \sum_i n_i^2 + V \sum_i P^{\dagger}_{00,i}
P_{00,i},
\label{hubbardS}
\end{equation}
with $U= 2 U_2$, $V = U_0 - U_2$, and $\tilde \mu=\mu- U/2$. 

When
$V=0$, i.e. $U_0 = U_2$, model (\ref{hubbardS}) corresponds to the
Hubbard model with an SU($2N$) spin symmetry: 
$c_{\alpha,i} \rightarrow \sum_{\beta} U_{\alpha\beta} c_{\beta,i}$,
$U$ being a SU($2N$) matrix.  The Hamiltonian (\ref{hubbardS}) for $V \ne
0$ still displays an extended symmetry.
Indeed,  since the BCS singlet-pairing operator (\ref{bcs}) 
is invariant under the Sp($2N$) group which 
consists of unitary matrices $U$ that satisfy
$U^* {\cal J} U^{\dagger} = {\cal J}$, model (\ref{hubbardS}) 
acquires an Sp($2N$) symmetry \cite{wusp}.  
In the spin $F=1/2$ case, model (\ref{hubbardS}) reduces to the SU(2) Hubbard
chain since SU(2) $\simeq$ Sp(2).  
In the $F=3/2$ case, i.e. $N=2$, there is no fine-tuning 
and models (\ref{hubbardSgen}) and (\ref{hubbardS}) have thus an exact
Sp(4)$\simeq$ SO(5) symmetry \cite{zhang}.
For $F > 3/2$, the Sp($2N$) symmetry is only present for the two coupling-constant model (\ref{hubbardS}). 
The Sp($2N$) symmetry simplifies the problem but may appear
rather artificial at this point.
However, one reason to  consider this model 
stems from the structure 
of the interaction involving the density $n_i$
and the BCS singlet-pairing term (\ref{bcs}) 
which are independent for $F \ge 3/2$
\footnote{ For $F=1/2$, one has indeed $P^\dagger_{00,i}P_{00,i}=n_i^2-n_i$.}.
The competition between these two operators gives rise to 
interesting physics. 
Indeed, at $U=0$ and large $V < 0$, 
a quasi-long-range BCS phase, which is a singlet
under Sp($2N$), should emerge. In contrast, for $V=0$ and $U <0$, 
we expect the stabilization of a charge-density wave (CDW) 
or a MS instability made of $2N$ fermions
$M_i = c_{1,i}^{\dagger} \ldots c_{2N,i}^{\dagger}$
which is an SU($2N$) singlet. 
We thus expect the existence of several 
distinct phases in the Sp($2N$) model (\ref{hubbardS}). 
As it will be revealed in the following,
the competition between these instabilities
relies on the existence of a ${\mathbb{Z}}_N$ symmetry which
is \emph{also} a symmetry of the original SU(2) model (\ref{hubbardSgen}).  
This ${\mathbb{Z}}_N$ symmetry can be defined by considering the coset
between the center ${\mathbb{Z}}_{2N}$ of the SU($2N$) group 
\footnote{We recall that the center of a 
group $G$ is defined from the elements which commute
with all elements of $G$. In particular, 
the Sp($2N$) or SU(2) groups share the same center
which is the ${\mathbb{Z}}_2$ group.}
and the center ${\mathbb{Z}}_2$ of the Sp($2N$) or SU(2):
${\mathbb{Z}}_{N}$ = ${\mathbb{Z}}_{2N}$/${\mathbb{Z}}_2$ with
\begin{eqnarray}
{\mathbb{Z}}_{2N} : c_{\alpha,i}^{\dagger} &\rightarrow& \mbox{e}^{in\pi/N}
c_{\alpha,i}^{\dagger} \; , n=0, ....,2N-1
\label{Z2Nsymmetry}
\\
{\mathbb{Z}}_{N} : c_{\alpha,i}^{\dagger} &\rightarrow& \mbox{e}^{in\pi/N}
c_{\alpha,i}^{\dagger} \; , n=0, ....,N-1,
\label{ZNsymmetry}
\end{eqnarray}
the ${\mathbb{Z}}_2$ symmetry being $c_{\alpha,i} \rightarrow -
c_{\alpha,i}$.  
The ${\mathbb{Z}}_N$ symmetry (\ref{ZNsymmetry}) provides an important
physical ingredient which is not present in the $F=1/2$ case
and stems from the higher-spin degeneracy.
The stabilization of a quasi-long-range BCS phase for $F>1/2$
requires the spontaneous breaking of this ${\mathbb{Z}}_N$ symmetry since the
singlet pairing $P^{\dagger}_{00,i}$ (\ref{bcs}) 
is \emph{not} invariant under this
symmetry.  
In contrast, if ${\mathbb{Z}}_N$ is not
broken, a MS phase,
which is a singlet under the ${\mathbb{Z}}_N$ symmetry, might emerge. 
In the following, the delicate competition between these
superfluid instabilities will be investigated 
with a special emphasis on this ${\mathbb{Z}}_N$ symmetry.
In particular, in the low-energy approach, this discrete symmetry 
is described by the ${\mathbb{Z}}_N$ parafermionic conformal field 
theory (CFT) which captures the universal properties 
of two-dimensional ${\mathbb{Z}}_N$ generalized Ising models \cite{para}.
This approach will enable us to determine the main features 
of the zero temperature phase diagram of model (\ref{hubbardS}).
A brief summary of our main results have already been published
as a letter \cite{phle}.

The rest of the paper is organized as follows.
The low-energy effective theory corresponding to 
model (\ref{hubbardS}) is developed in
section II.
The nature of the different phases is then deduced 
from  a  renormalization group (RG) analysis. 
In section III, we present the phase diagram
of the model (\ref{hubbardS}) at zero temperature 
for incommensurate filling and
we discuss the main physical properties of the phases as well 
as the nature of the quantum phase transitions.
The Mott-insulating phases for the case of one atom per site 
are also investigated.
Finally, we conclude in section IV
and some technical details are presented in three appendices.

\section{Low-energy approach}

In this section, we present the low-energy description 
of the spin-$F$ Hubbard model with 
a Sp($2N$) symmetry (\ref{hubbardS})  which will 
enable us to determine its phase diagram 
at zero temperature in the next section.
 
\subsection{Continuum limit}

Let us first discuss the continuum limit of the lattice
model (\ref{hubbardS}).
Its low-energy effective field theory can be derived from the continuum
description of the lattice fermionic operators $c_{\alpha,i}$ in
terms of right and left-moving Dirac fermions (for a review, see for 
instance the books \cite{tsvelikbook,bookboso,giamarchi}):
\begin{equation}
c_{\alpha,i}/\sqrt{a_0} \rightarrow R_{\alpha}
\mbox{e}^{ik_F x} + L_{\alpha} \mbox{e}^{-ik_F x},
\label{contfer}
\end{equation}
with $x= i a_0$ ($a_0$ being the lattice spacing) and $k_F$ is the
Fermi momentum.  
The left (right)-moving fermions are holomorphic (antiholomorphic) 
fields of the complex coordinate $z= v_F\tau + ix$ ($\tau$ being
the imaginary time and $v_F$ is the Fermi velocity): 
$R_{\alpha}(\bar z), L_{\alpha}(z)$. 
These fields  obey the following operator product expansion (OPE):
\begin{eqnarray}
R_{\alpha}^{\dagger} \left(\bar z \right)  
R_{\beta} \left(\bar \omega \right) &\sim&
\frac{\delta_{\alpha \beta}}{2 \pi \left( \bar z - \bar \omega \right)} 
\nonumber \\
L_{\alpha}^{\dagger} \left(z \right)  L_{\beta} \left( \omega \right) &\sim&
\frac{\delta_{\alpha \beta}}{2 \pi \left(z - \omega \right)} .
\label{OPEfer}
\end{eqnarray}
In this continuum limit, 
the non-interacting part of the Hamiltonian (\ref{hubbardS})
corresponds to the Hamiltonian density of 
$2N$ free relativistic massless fermions:
\begin{equation}
{\cal H}_0 = - i v_F \left( : R_{\alpha}^{\dagger} \partial_x R_{\alpha} :
- : L_{\alpha}^{\dagger} \partial_x L_{\alpha} : \right),
\label{contfreeham}
\end{equation}
where $v_F = 2t a_0 \sin (k_F a_0)$ and
the normal ordering $::$ with
respect to the Fermi sea is assumed as well as a summation over 
repeated indices.
The continuous symmetry of the non-interacting 
part of the Hamiltonian (\ref{hubbardS}) 
is enlarged, in the continuum limit, to U(2$N$)$|_L$ $\otimes$ U(2$N$)$|_R$
under independent unitary transformations on the Dirac fermions. 

The next step of the approach is to use the non-Abelian
bosonization \cite{witten,knizhnik,affleckspinchain} to 
investigate the effect of the 
Sp(2$N$) symmetry of the lattice model.
To this end, we introduce a SU(2$N$)$_1$ 
Wess-Zumino-Novikov-Witten (WZNW) primary
field $g_{\alpha \beta}$ to represent the spin degrees of freedom 
and an additional U(1) charge boson field $\Phi_c$. 
The free-fermion theory (\ref{contfreeham}) is then equivalent
to the CFT with an 
U(1) $\times$ SU(2$N$)$_1$ symmetry.
The right and left-moving fermions can be written 
as a product of spin and charge operators: 
\begin{eqnarray}
R_{\alpha} &\sim&  :\exp\left(i \sqrt{2 \pi/N} \; \Phi_{cR} \right): g_{\alpha R}
\nonumber \\
L_{\alpha}^{\dagger} &\sim&  
:\exp\left(i \sqrt{2 \pi/N} \; \Phi_{cL} \right): g_{\alpha L},
\label{fernonabelboso}
\end{eqnarray}
where $g_{R}$ and $g_L$ are respectively 
the right and left parts of the SU(2$N$)$_1$
 primary field $g$ with scaling dimension $(2N- 1)/2N$ 
which transforms in the fundamental 
representation of SU(2$N$).
In Eq. (\ref{fernonabelboso}), $\Phi_{cR,L}$ denote the chiral parts
of the charge boson field: $\Phi_{c} = \Phi_{cR} + \Phi_{cL}$.
As is well known, the free-Hamiltonian (\ref{contfreeham}) 
can be decomposed into charge and spin 
pieces:
\begin{equation}
{\cal H}_0 = \frac{v_F}{2}
\left[
\left(\partial_x \Phi_c \right)^2 + 
\left(\partial_x \Theta_c \right)^2 \right]
+ \frac{2\pi v_F}{2N + 1} \left[ : I^A_{R} I^A_{R}: + : I^A_{L} I^A_{L}: 
\right],
\label{contfreehambis}
\end{equation}
where $\Theta_c = \Phi_{cL} - \Phi_{cR}$ is the dual charge boson field
and $I^A_{R,L} (A=1,\ldots,4 N^2 -1)$ 
are the currents which generate the SU(2$N$)$_1$ conformal 
symmetry with central charge
$c= 2N -1$.
They express simply as bilinears of the Dirac
fermions: 
\begin{equation}
I^A_{R} = :R_{\alpha}^{\dagger} T^A_{\alpha \beta}
R_{\beta}: , \; \;  
I^A_{L} = :L_{\alpha}^{\dagger} T^A_{\alpha \beta}
L_{\beta}: ,
\label{su2Ncur}
\end{equation}
$T^A$ being the generators of SU($2N$) in the fundamental
representation normalized such that: 
$\mbox{Tr} (T^A T^B) = \delta^{AB}/2$ (see Appendix A).  
These currents satisfy the SU(2$N$)$_1$ current algebra \cite{bookboso,dms}:
\begin{eqnarray}
I_L^A\left(z\right) I_L^B\left(\omega\right) &\sim& \frac{\delta^{AB}}{8 \pi^2 
\left(z - \omega \right)^2} + 
\frac{i f^{ABC}}{2 \pi \left(z - \omega \right)} I_L^C\left(\omega\right)
\nonumber \\
I_R^A\left(\bar z\right) I_R^B\left(\bar \omega\right) 
&\sim& \frac{\delta^{AB}}{8 \pi^2
\left(\bar z - \bar \omega \right)^2} +
\frac{i f^{ABC}}{2 \pi \left(\bar z - \bar \omega \right)}
I_R^C\left(\bar \omega\right),
\label{curralg}
\end{eqnarray}
where $f^{ABC}$ are the structure constants of the SU(2$N$) group.

In the presence of interactions, the spin-charge separation  
(\ref{contfreehambis})  still holds away from half-filling
(i.e. $N$ atoms per site). 
In that case, the low-energy Hamiltonian of
model (\ref{hubbardS}) separates into two commuting charge and
spin pieces:
\begin{equation}
{\cal H} = {\cal H}_c + {\cal H}_s, \; \;[{\cal H}_c, {\cal H}_s] = 0.
\label{spinchargesepar}
\end{equation}
Let us first consider the charge degrees of freedom which 
are described by the bosonic field $\Phi_c$.

\subsubsection{Charge degrees of freedom}
Using the continuum description (\ref{contfer}) and the 
decomposition (\ref{fernonabelboso}), we find 
that the charge excitations of model (\ref{hubbardS})
are captured by the following Hamiltonian: 
\begin{equation}
{\cal H}_c = \frac{v_F}{2} \left[
\left(\partial_x \Phi_c \right)^2 +
\left(\partial_x \Theta_c \right)^2 \right]
+ a_0\frac{2 V + U N\left(2N -1\right)}{2 N \pi}
\left(\partial_x \Phi_c \right)^2 .
\label{chargeham}
\end{equation}
It can be recast into the form of the Tomonaga-Luttinger 
free Hamiltonian:
\begin{equation}
{\cal H}_c = \frac{v_c}{2} \left[\frac{1}{K_c}
\left(\partial_x \Phi_c \right)^2 + K_c
\left(\partial_x \Theta_c \right)^2 \right],
\label{luttbis}
\end{equation}
where the charge velocity $v_c$ and the 
Luttinger parameter $K_c$ read as follows: 
\begin{eqnarray}
v_c &=& v_F \left[1 + a_0 (2 V + U N(2N -1))/(N\pi v_F)\right]^{1/2}
\nonumber \\
K_c &=& \left[1 + a_0 (2 V + U N(2N -1))/(N\pi v_F)\right]^{-1/2} .
\label{luttpara}
\end{eqnarray}
The conserved quantities in this U(1) charge sector
are the total particle number ${\cal N}$ and current ${\cal I}$: 
\begin{eqnarray}
{\cal N} &=& \int d x \;
:R^{\dagger}_{\alpha} R_{\alpha} + L^{\dagger}_{\alpha} L_{\alpha}: \; =
\sqrt{2N/\pi} \int d x \; \partial_x \Phi_c \nonumber \\
{\cal I} &=& \int d x \;
:L^{\dagger}_{\alpha} L_{\alpha} - R^{\dagger}_{\alpha} R_{\alpha}: =
\sqrt{2N/\pi} \int d x \; \partial_x \Theta_c .
\label{chargeconserquant}
\end{eqnarray}
For generic fillings, no umklapp terms appear and the charge
degrees of freedom display metallic (gapless) properties in the
Luttinger liquid universality class \cite{bookboso,giamarchi}.
However, as it will be discussed in section III,
the existence of an umklapp process, for a commensurate filling of
one atom per site, is responsible for the formation of  
a charge gap and the emergence of 
different Mott insulating phases. 

\subsubsection{Spin-degrees of freedom}
All non-trivial physics corresponding to the spin degeneracy 
is encoded in the spin part ${\cal H}_s$
of the Hamiltonian (\ref{spinchargesepar}). 
Its continuum expression can be obtained by decomposing 
the SU($2N$)$_1$ currents of Eq. (\ref{su2Ncur}) into
$\parallel$ and $\perp$
parts $I^A_{R,L}= (I^a_{\parallel}, I^i_{\perp})_{R,L}$ with respect to the
Sp($2N$) symmetry of the lattice model (\ref{hubbardS}).  The currents
$I^a_{\parallel R(L)}, a=1,..,N(2N+1)$ generate the Sp($2N$)$_1$ CFT
symmetry with central charge $c=N(2N+1)/(N+2)$. 
They can be simply expressed in terms of the chiral Dirac
fermions: 
\begin{equation}
I^a_{\parallel R} = :R_{\alpha}^{\dagger} T^a_{\alpha \beta}
R_{\beta}: , \; \;
I^a_{\parallel L} = :L_{\alpha}^{\dagger} T^a_{\alpha \beta}
L_{\beta}: ,
\label{sp2ncur}
\end{equation}
where $T^a$ are the generators of Sp($2N$) in the fundamental
representation and normalized 
such that: $\mbox{Tr} (T^a T^b) = \delta^{ab}/2$ (see Appendix A).  
These currents verify the Sp($2N$)$_1$ Kac-Moody algebra
given by Eqs. (\ref{curralg}) with $f^{abc}$ the Sp($2N$) structure constants.
The remaining SU($2N$)$_1$ 
currents $I^i_{\perp R,L} $ ($i=1,..,2N^2 - N - 1$) are written as:
\begin{equation}
I^i_{\perp R} = :R_{\alpha}^{\dagger} T^i_{\alpha \beta}
R_{\beta}:, \; \;  I^i_{\perp L} = :L_{\alpha}^{\dagger} T^i_{\alpha \beta}
L_{\beta}: .
\label{wrongcur}
\end{equation}

With these definitions, we can now derive the continuum description 
of the spin degrees of freedom using Eq. (\ref{contfer})
and Eqs. (\ref{sumsu2ngen}, \ref{sumsp2ngen}) of Appendix A.
After some cumbersome calculations, we find that the low-energy
Hamiltonian in the spin sector, ${\cal H}_s$, can be expressed in terms of the currents only, and displays a marginal current-current interaction:
\begin{equation}
{\cal H}_s = 
\frac{2\pi v_{s\parallel} }{2N + 1} \left[ :I^a_{\parallel R}
I^a_{\parallel R}: + :I^a_{\parallel L} I^a_{\parallel L}:
\right] + 
\frac{2\pi v_{s\perp} }{2N + 1} \left[
:I^i_{\perp R} I^i_{\perp R}: + :I^i_{\perp L} I^i_{\perp L}:
\right] + g_{\parallel} I^a_{\parallel R}
I^a_{\parallel L} + g_{\perp} I^i_{\perp R} I^i_{\perp L},
\label{spinham}
\end{equation}
with $g_{\parallel} = - 2 a_0 (2V + N U)/N$, 
$g_{\perp} = 2 a_0 (2V - N U)/N$ and $v_{s\parallel, \perp}$ 
are the spin velocities:
$v_{s\parallel} = v_F - U a_0/2\pi - V a_0/\pi$, 
$v_{s\perp} = v_F - U a_0/2\pi + V a_0(N+1)/N\pi$.
In the simplest $N=1$ case, we have  Sp(2)$_1$ $\sim$ SU(2)$_1$ 
and the Hamiltonian (\ref{spinham}) with $g_{\perp} = 0$ 
describes the spin-sector of the continuum limit of the spin-1/2 Hubbard chain 
which can be found for instance in the book \cite{bookboso} written 
with the same notations used in this work.
In the $N=2$, i.e. the $F=3/2$ case, one can express model (\ref{spinham})
in a more transparent basis.
Indeed, there is a simple free-field
representation of the 
unperturbed SU(4)$_1$ $\sim$ SO(6)$_1$ CFT in terms of six real (Majorana)
fermions which has been used in the context of the two-leg
spin-1/2 ladder with four-spin exchange interactions \cite{aza4spin}. 
Introducing six real fermions $\xi^{0}_{R,L}$ and
$\xi^{i}_{R,L} ,i=1,...,5$ to describe respectively the
${\mathbb{Z}}_2$, i.e. Ising, and SO(5)$_1$ $\simeq$ Sp(4)$_1$ CFTs, 
the currents of Eq. (\ref{spinham})
can be written locally in terms of these fermions:
\begin{eqnarray}
I^i_{\perp R,L} &\sim& - \frac{i}{\sqrt{2}}  \xi^{0}_{R,L} \xi^{i}_{R,L}  
\nonumber \\
I^a_{\parallel R,L} &\sim& - \frac{i}{\sqrt{2}}   \xi^{i}_{R,L} \xi^{j}_{R,L}  ,
\label{freeferep}
\end{eqnarray}
where $a=1, \ldots, 10 \equiv (i,j)$ 
where $1 \le i<j \le 5$.
The Majorana fermions are normalized as 
the Dirac fermions of Eq. (\ref{OPEfer}) to reproduce
faithfully the Kac-Moody algebra (\ref{curralg}).
The current-current model 
of Eq. (\ref{spinham}) can then be expressed in terms 
of these real fermions:
\begin{equation}
{\cal H}_s = -\frac{i v}{2} \left[:\xi^{i}_R \partial_x 
\xi^{i}_R: - :\xi^{i}_L \partial_x \xi^{i}_L: \right]
- \frac{i v_0}{2} \left[:\xi^{0}_R \partial_x
\xi^{0}_R: - :\xi^{0}_L \partial_x \xi^{0}_L: \right]
+ \lambda_{\parallel} (\xi^{i}_R \xi^{i}_L)^2 + \lambda_{\perp}
\xi^{0}_{R} \xi^{0}_{L} \xi^{i}_R \xi^{i}_L ,
\label{spin3demirefer}
\end{equation}
with $\lambda_{\parallel} = - a_0 (U + V)$, 
$\lambda_{\perp} = a_0 (V - U)$, and 
the spin velocities:
$v = v_F - a_0(U +V)/2 \pi$, 
$v_0 = v_F - a_0(U - 3 V)/2 \pi$.
In absence of the spin-velocity anisotropy, model (\ref{spin3demirefer})
turns out to be exactly solvable and has also
been studied in the context of a 
SO(5) symmetric two-leg ladder 
\cite{controzzi}.

\subsection{RG analysis}
\label{RGanalysis}
In the general $N$ case, model (\ref{spinham}) is not
integrable and  the main effect of the current-current interaction 
can be elucidated by means of a RG analysis.
The one-loop RG equations of model (\ref{spinham})
are given by (see Appendix B):
\begin{eqnarray}
{\dot g}_{\parallel} &=&
\frac{N + 1}{4 \pi} \; g_{\parallel}^2
+ \frac{N - 1}{4 \pi} \; g_{\perp}^2 \nonumber \\
{\dot g}_{\perp} &=& \frac{N}{2 \pi} \; g_{\perp} g_{\parallel} ,
\label{1RGeqs}
\end{eqnarray}
where ${\dot g}_{\perp,\parallel} = \partial g_{\perp,\parallel}/\partial t$,
$t$ being the RG time.
In Eq. (\ref{1RGeqs}), we have neglected the spin-velocity anisotropy 
$v_s = v_{s \parallel} \simeq v_{s \perp}$ and have absorbed 
$v_s$ in a redefinition of the coupling constants: $g_\alpha\rightarrow g_\alpha/v_s$.
We have also obtained the two-loop RG equations of model (\ref{spinham})
and the results are presented in Appendix B.
The one-loop RG equations (\ref{1RGeqs}) 
can be solved and in particular the RG invariant
flow $K$ which parametrizes the RG lines reads as follows:
\begin{equation}
K = g_{\perp}^{-(N+1)/N} \left(g_{\parallel}^2 - g_{\perp}^2 \right).
\label{invflot}
\end{equation}
The RG flow emerging from Eqs. (\ref{1RGeqs}) is rich and
consists of three different phases (see Fig. 1).
\begin{figure}
\begin{center}
\includegraphics[width=0.8\linewidth]{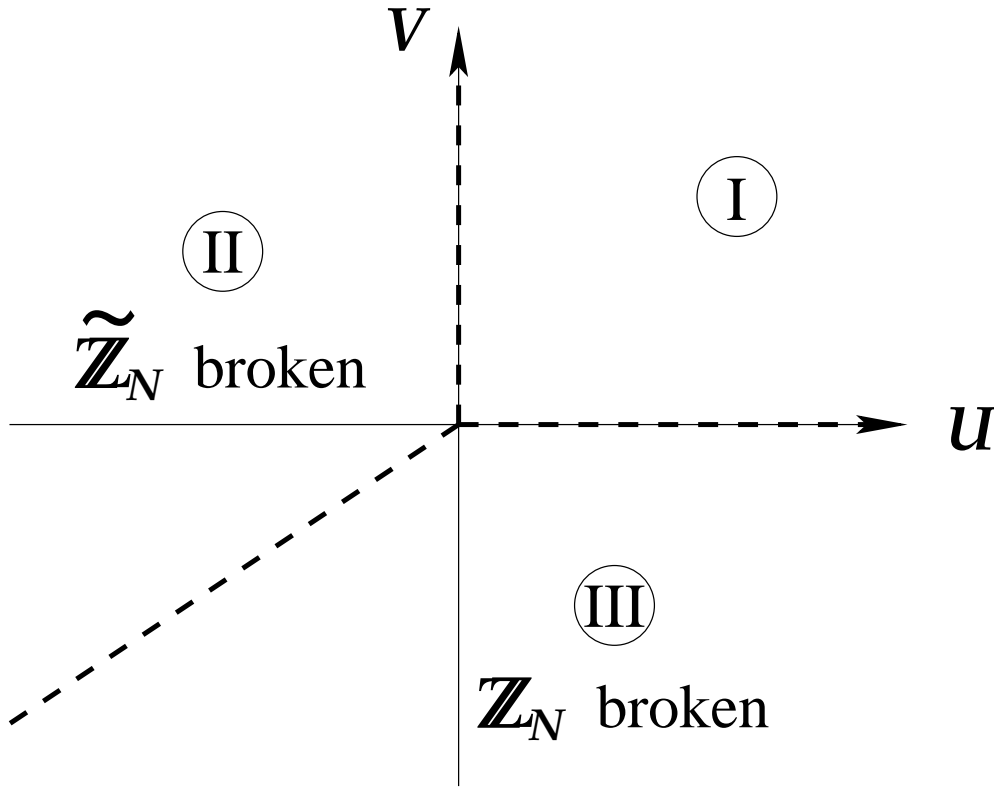}
\end{center}
\caption{
Phase diagram of model (\ref{spinham}) obtained from the RG approach.
Phase I is gapless while phases II and III have gapped spin excitations.
The dash lines stand for the phase transition lines 
predicted from the RG calculation.
}
\end{figure}

In region I, where both $U$ and $V$ are positive,
all couplings go to zero in the infrared (IR) limit
and the interaction is marginal irrelevant.
The symmetry of the IR fixed point is
SU(2$N$)$_1$ (up to a velocity anisotropy) leading to
$2N -1$ gapless spin excitations.
The current-current interaction of Eq. (\ref{spinham}) leads to logarithmic 
corrections to physical quantities \cite{Affleck-G-S-Z-89,majumdar}.
The low-energy properties of this phase are very similar
to that of the repulsive SU(2$N$) Hubbard chain
which have been studied in Refs. \cite{affleck,assaraf}.
In contrast,
a spin gap is opened by the interaction
in the two remaining phases.
In phase II, defined by $U < 0$ and
$V > N U /2$, the RG flow in the far IR limit is
attracted along a special symmetric ray $ g_{\parallel} = g_{\perp} =
g^{*} > 0$ where the interacting
part of the Hamiltonian (\ref{spinham})
can be rewritten in a manifest SU(2$N$) invariant form:
\begin{equation}
{\cal H}_{{\rm s},{\rm int}}^{*} =
g^{*}\left(
I^a_{\parallel R} I^a_{\parallel L}
+ I^i_{\perp R} I^i_{\perp L} \right)
= g^{*} I^A_R I^A_L .
\label{su2NGN}
\end{equation}
This IR Hamiltonian thus takes 
the form of the SU(2$N$) Gross-Neveu (GN) model \cite{gross} which
is an integrable massive field theory \cite{andrei}.
The development of the strong-coupling regime in
the SU(2$N$) GN model leads to the generation of a spin
gap. 
The low-energy properties of the spin
sector of phase II can be extracted from the integrability of the
SU($2N$) GN model (\ref{su2NGN}).  
Its low-energy spectrum consists into
$2N-1$ branches with masses: $m_r = m \sin(\pi r /2N)$
that transform in the SU(2$N$) representation with Young
tableau with one column and $r$ boxes ($r = 1, \ldots, 2N-1$)
\cite{andrei}.
The corresponding eigenstates are labelled by 
quantum numbers associated with the
conserved quantities of the SU($2N$) low-energy symmetry (the Cartan basis):
$Q_{\parallel}^a = \int dx \; (I^a_{\parallel L} + I^a_{\parallel
  R})$, $a= (1,..., N)$ and $Q_{\perp}^i = \int dx \; (I^i_{\perp L }
+ I^i_{\perp R })$, $i= (1,..., N-1)$.  Due to the Sp($2N$) symmetry
of model (\ref{hubbardS}), the $Q_{\parallel}^a$ numbers are conserved
whereas the $Q_{\perp}^i$ charges are only good quantum numbers at low
energy. 
This is an example of a dynamical symmetry enlargement which 
corresponds to a situation where a Hamiltonian is attracted 
under a RG flow to a manifold possessing a higher symmetry
than indicated by the original microscopic theory.
This phenomenon occurs in a large variety of models with marginal interactions 
in the scaling limit \cite{konik} as, for instance,
in the half-filled two-leg Hubbard model \cite{balents}
and in the  SU(4) Hubbard model at half filling \cite{boulat}, where a SO(8) symmetry emerges at low energy.

In the second spin-gapped phase (III) of Fig. 1, defined by
$V<0$ and $V < N U/2$, the RG flow is now attracted along the
asymptote: $g_{\parallel} = - g_{\perp} = g^{*} > 0$.  In that case,
the interacting part of the IR Hamiltonian becomes
\begin{equation}
{\cal H}_{{\rm s},{\rm int}}^{*} = g^{*}\left( I^a_{\parallel R}
I^a_{\parallel L} - I^i_{\perp R} I^i_{\perp L} \right),
\label{su2NGNdual}
\end{equation}
which can be recast as a SU($2N$) GN model (\ref{su2NGN}) by means of
a \emph{duality} transformation ${\cal D}$ on the fermions: ${\cal
D}R(L){\cal D}^{-1} = {\tilde R(\tilde {L})}$ with 
\begin{equation}
{\tilde
R}_{\alpha} = {\cal J}_{\alpha \beta} R^{\dagger}_{\beta}, \; \;
{\tilde L}_{\alpha}= L_{\alpha}.  
\label{dualityfer}
\end{equation}
Using Eqs. (\ref{sp2ncur}, \ref{wrongcur}), we observe that
this transformation acts on the
currents as: ${\tilde I}^a_{\parallel R(L)} = I^a_{\parallel R(L)}$
and $ {\tilde I}^i_{\perp R(L)}= - (+) I^i_{\perp R(L)}$ so that
${\cal D}$ indeed maps (\ref{su2NGNdual}) onto (\ref{su2NGN}).
Besides the opening of a spectral gap, we thus find that phase III
possesses a hidden symmetry at low energy i.e.  a $\widetilde{SU}(2N)$
symmetry generated by the dual currents $({\tilde I}^a_{\parallel
R(L)},{\tilde I}^i_{\perp R(L)})$.  
The spin spectrum in phase III can be obtained from the
duality symmetry ${\cal D}$ and consists into the  $2N -1$ branches 
$m_r$ which
transform in the representations of the dual group
$\widetilde{SU}(2N)$.  The dual quantum numbers are now given by:
${\tilde Q}_{\parallel}^a = Q_{\parallel}^a$ and ${\tilde Q}_{\perp}^i
= \int dx \; ({\tilde I}^i_{\perp L } + {\tilde I}^i_{\perp R }) =
\int dx \; (I^i_{\perp L } - I^i_{\perp R }) = {\cal I}_{\perp}^i$.  We thus
observe that the low-lying excitations in phase III carry quantized
spin \emph{currents} in the ``$\perp$'' direction.  In this sense, the
phase III might be viewed as a partially spin-superfluid phase.
In summary, the existence of these two distinct spin-gapped phases is a
non-trivial consequence of higher-spin degeneracy and does not occur
in the $F=1/2$ case.  

\subsection{Conformal embedding}

The crucial point of a non-perturbative analysis is 
often the identification of a good
basis that describe low-lying excitations of the phase.
Much insight on this problem can be gained from the symmetries
of the model and the use of the
non-Abelian bosonization approach for 1D systems.
Such an approach has been extremely powerful in the past
as in quantum impurity
problems \cite{affleckondo} and 
in spin chains \cite{affleckspinchain,bookboso}.
So far, we have used an U(2$N$)$_1$ $=$ U(1) $\times$
SU(2$N$)$_1$ CFT approach  to determine the low-energy 
properties of model (\ref{hubbardS}).
However, this description is not adequate to 
give a full understanding of the two spin-gapped phases
found in the RG approach. 
In particular, the physical origin of the formation of 
the spin gaps is not clear at this point.
What is the nature of the discrete symmetry 
which is spontaneously broken in phases II and III?
A second weak point of the previous analysis is the 
determination of the quantum phase transition
between the two spin-gapped phases which is rather unclear 
within the preceding description. At least,
this transition should occur in the manifold which
is invariant under the duality ${\cal D}$ symmetry (\ref{dualityfer}) i.e. 
the self-dual manifold defined by: $g_{\perp} = 0$. 
The low-energy Hamiltonian which describes the phase transition
is thus (neglecting the spin-velocity anisotropy):
\begin{equation}
{\cal H}_s^{\rm SD} = \frac{2\pi v_s}{2N + 1} \left[ :I^a_{\parallel R}
I^a_{\parallel R}: +  :I^a_{\parallel L}
I^a_{\parallel L}: +
:I^i_{\perp R} I^i_{\perp R}: 
+ :I^i_{\perp L} I^i_{\perp L}: \right] + g_{\parallel}^{\rm SD} I^a_{\parallel R}
I^a_{\parallel L} ,
\label{spinhamtrans}
\end{equation}
with $g_{\parallel}^{\rm SD} >0$ so that the marginally relevant current-current interaction
opens a mass gap in the Sp(2$N$) sector. However, one cannot conclude on the 
occurrence of the first-order phase transition since the non-interacting
Hamiltonian of Eq. (\ref{spinhamtrans}) contains more degrees of freedom
than the  Sp(2$N$) one. 
They remain massless and control 
the quantum phase transition.
The nature of these decoupled degrees of freedom is not clear at this point.

Prompted by all these questions, it is important to fully exploit 
the existence of the Sp(2$N$) symmetry of the lattice model (\ref{hubbardS})
and to consider the following conformal embedding:
U(2$N$)$_1$ $\rightarrow$ U(1) $\times$ Sp(2$N$)$_1$ $\times$
[SU(2$N$)$_1$/Sp(2$N$)$_1$].
The coset SU(2$N$)$_1$/Sp(2$N$)$_1$ CFT has central
charge $c= 2N - 1 - N(2N+1)/(N+2) = 2(N-1)/(N+2)$
which is that of the ${\mathbb{Z}}_N$ parafermionic CFT \cite{para}.
In fact, as shown by Altschuler \cite{altschuler}, it turns out that
the SU(2N)$_1$/Sp(2N)$_1$ CFT
is indeed equivalent to the 
${\mathbb{Z}}_N$ parafermionic CFT
which describes self-dual critical points of two-dimensional ${\mathbb{Z}}_N$ 
Ising models (see Appendix C).
The conformal embedding U(2$N$)$_1$ $\rightarrow$ 
U(1) $\times$ Sp(2$N$)$_1$ $\times$ 
${\mathbb{Z}}_N$ provides
us with a \emph{non-perturbative} basis to express
any physical operator in terms
of its charge and spin degrees of freedom which are described
respectively by
the U(1) and Sp(2$N$)$_1$ $\times$ ${\mathbb{Z}}_N$ CFTs.
Since Sp(2) $\sim$ SU(2),
this basis for $N=1$ accounts for the well-known spin-charge
separation which is the hallmark
of 1D spin-1/2 electronic systems \cite{bookboso,giamarchi}.
In this respect, in the general half-odd integer spin case,
the ${\mathbb{Z}}_N$ symmetry plays its trick and provides a new
important ingredient not present for $F=1/2$.
In the low-energy approach, the spin degrees of freedom
corresponding to this symmetry are captured by an effective
2D ${\mathbb{Z}}_N$ Ising model.
As in the $N=2$
case, these ${\mathbb{Z}}_N$ Ising models exhibit two gapped phases described
by order and disorder parameters $\sigma_k$ and $\mu_k$ ($k=1,..,N-1$)
which are dual to each other by means of the Kramers-Wannier (KW)
duality symmetry.  This duality transformation maps the
${\mathbb{Z}}_N$ symmetry, spontaneously broken in the low-temperature phase
($\langle \sigma_k \rangle \ne 0$ and $\langle \mu_k \rangle = 0$),
onto a ${\tilde {\mathbb{Z}}}_N$ symmetry which is broken in the
high-temperature phase where $\langle \mu_k \rangle \ne 0$ and
$\langle \sigma_k \rangle = 0$.  At the critical point, the theory is
self-dual with a ${\mathbb{Z}}_N$ $\times$ ${\tilde {\mathbb{Z}}}_N$
symmetry and its universal properties are captured by the
${\mathbb{Z}}_N$ parafermionic CFT with
$\sigma_k, \mu_k$ becoming conformal fields with scaling
dimension $d_k = k(N-k)/N(N+2)$ \cite{para}.
This ${\mathbb{Z}}_N$ CFT is generated by 
right and left parafermionic currents $\Psi_{k R,L}$ ($\Psi_{k R,L}^{\dagger} 
= \Psi_{N-k R,L}$, $k=1,\ldots, N-1$)
with scaling dimension $\Delta_k = k(N-k)/N$ 
which are the generalization of the Majorana fermions of the ${\mathbb{Z}}_2$ Ising model.
Under the ${\mathbb{Z}}_N$ $\times$ ${\tilde {\mathbb{Z}}}_N$ symmetry, 
$\Psi_{k L}$ (respectively $\Psi_{k R}$) carries
a $(k,k)$ (respectively $(k,-k)$) charge which means:
\begin{eqnarray}
\Psi_{k L,R} &\rightarrow& e^{i 2 \pi m k/N} \Psi_{k L,R} \; \; {\rm under} \; \; \mathbb{Z}_N 
\nonumber \\ 
\Psi_{k L,R} &\rightarrow& e^{\pm i 2 \pi m k/N} \Psi_{k L,R} 
\; \; {\rm under} \; \; {\tilde {\mathbb Z}}_N  ,
\label{chargepara}
\end{eqnarray}
with $m=0,\ldots, N-1$.
As it is discussed in Appendix C, there is a faithful representation
of the first parafermionic currents $\Psi_{1 L,R}$ in terms of the Dirac fermions and 
the charge bosonic field $\Phi_{cR,L}$:
\begin{eqnarray}
L_{\alpha}^{\dagger} {\cal J}_{\alpha \beta} L_{\beta}^{\dagger}  
&\simeq& \frac{\sqrt{N}}{\pi} 
:\exp \left( i \sqrt{8 \pi/N} \; \Phi_{cL} \right): \Psi_{1 L} 
\nonumber \\
R_{\alpha}^{\dagger} {\cal J}_{\alpha \beta} R_{\beta}^{\dagger}
&\simeq& \frac{\sqrt{N}}{\pi} 
:\exp \left( -i \sqrt{8 \pi/N} \; \Phi_{cR} \right): \Psi_{1 R} .
\label{pararep}
\end{eqnarray}
We deduce, from this identification, that the duality 
symmetry (\ref{dualityfer}) of the RG analysis together
with the Gaussian duality ($\Phi_{c R,L} \rightarrow \mp \Phi_{c R,L}$) 
give $\Psi_{1 L} \rightarrow  \Psi_{1 L}$
and $\Psi_{1 R} \rightarrow - \Psi_{1 R}^{\dagger}$
which is nothing but the KW duality transformation on the first parafermionic current 
(see Eq. (\ref{KWparafer}) of Appendix C). 
It is thus tempting to interpret the formation of the spin-gapped phases of Fig. 1
as the result of the spontaneous breaking of the $\mathbb{Z}_N$ 
and ${\tilde {\mathbb Z}}_N$ discrete symmetries.
In this respect, from the correspondence (\ref{pararep})
and Eq. (\ref{chargepara}), we observe that under the $\mathbb{Z}_N$ symmetry,
the left and right-moving fermions transform as:
\begin{equation}
L_{\alpha} \rightarrow e^{-i \pi m /N} L_{\alpha}, \; \;
R_{\alpha} \rightarrow e^{-i \pi m /N} R_{\alpha} ,
\label{znfer}
\end{equation}
while under ${\tilde {\mathbb Z}}_N$ we have:
\begin{equation}
L_{\alpha} \rightarrow e^{-i \pi m /N} L_{\alpha}, \; \;
R_{\alpha} \rightarrow e^{i \pi m /N} R_{\alpha} .
\label{tildeznfer}
\end{equation}
Using the continuum representation of the fermions (\ref{contfer}), we find that
the lattice ${\mathbb{Z}}_N$ symmetry (\ref{ZNsymmetry}) corresponds to 
the ${\mathbb{Z}}_N$ symmetry of
an effective 2D ${\mathbb{Z}}_N$ Ising model. 
In contrast, the ${\tilde {\mathbb Z}}_N$ symmetry has no simple local lattice
representation.  

We need now to fully identify phases II and III of Fig. 1 with
the high and low-temperature phases of the ${\mathbb{Z}}_N$ Ising model.
To this end, we use the fact that
the SU($2N$)$_1$ WZNW primary field $g$ of Eq. (\ref{fernonabelboso})  can be expressed in terms 
of the Sp(2$N$)$_1$ $\times$ ${\mathbb{Z}}_N$ basis 
using the conformal embedding (see Eq. (\ref{grepapp}) of Appendix C): 
\begin{equation}
:\exp \left(-i\sqrt{2 \pi/N} \Phi_c\right): L^{\dagger}_{\alpha} R_{\alpha} 
 \sim {\rm Tr} \; g  \sim \mu_1 {\rm Tr} \; \phi^{(1)},
\label{grepzn}
\end{equation}
where $\phi^{(1)}$ is the Sp($2N$)$_1$ primary field 
with scaling dimension $\Delta = (2N+1)/2(N+2)$. 
We deduce, from the identification (\ref{grepzn}), that
${\rm Tr} \; g$ has the same behavior as $\mu_1$ under the 
${\mathbb{Z}}_N \times {\tilde {\mathbb Z}}_N$ symmetry 
(Eq. (\ref{chargeorderdisorder}) of Appendix C):
\begin{eqnarray}
{\rm Tr} \; g  &\rightarrow& {\rm Tr} \; g \; \; {\rm under} \; \; \mathbb{Z}_N
\nonumber \\
{\rm Tr} \; g &\rightarrow& e^{i 2 \pi m /N} {\rm Tr} \; g 
\; \; {\rm under} \; \; {\tilde {\mathbb Z}}_N  .
\label{trgsym}
\end{eqnarray}
The next step of the approach is to note that
the interacting part of the Hamiltonian (\ref{su2NGN}), which controls
the strong coupling behavior of the RG flow in phase II, can be expressed 
in terms of ${\rm Tr} \; g$: 
${\cal H}^{*}_{s, {\rm int}} \sim |{\rm Tr} \; g |^2$.
The ground state of this phase displays long-range order associated
with the order parameter ${\rm Tr} \; g$: $\langle {\rm Tr} \; g \rangle \ne 0$.
According to Eq. (\ref{trgsym}), we then deduce that 
the ${\tilde {\mathbb Z}}_N$ symmetry is spontaneously broken 
while the $\mathbb{Z}_N$ symmetry remains unbroken in phase II.
The $\mathbb{Z}_N$ Ising model thus belongs  to its high-temperature phase
and a spectral gap is formed. Using the KW duality symmetry or 
the transformation (\ref{dualityfer}) on the Dirac fermions, 
one can conclude that phase III corresponds to the low-temperature phase of 
the $\mathbb{Z}_N$ Ising model where the $\mathbb{Z}_N$ symmetry 
is spontaneously broken. 

In summary,
the existence of the two spin-gapped phases of Fig. 1
is a non-trivial consequence of higher-spin degeneracy
and does not occur in the $F=1/2$ case.
The emergence of the spin-gap stems from
the spontaneous breakdown of the ${\mathbb{Z}}_N$ or ${\tilde {\mathbb{Z}}}_N$
discrete symmetries.
As we shall see now, these symmetries are central to
the striking physical properties displayed by
these phases.

\section{Phase diagram}

In this section, we discuss the phase diagram at zero temperature
of the lattice model (\ref{hubbardS}) for incommensurate
filling and for a commensurate filling of one atom per site.
The nature of the quantum phase transitions will also be investigated.
Let us start with the incommensurate filling case.

\subsection{Phase diagram for incommensurate filling}

We shall now determine the nature of the dominant
electronic instabilities of the different phases of Fig. 1
for incommensurate filling.  
The RG analysis of the preceding section reveals the existence
of three different phases.

\begin{figure}
\begin{center}
\includegraphics[width=0.8\linewidth]{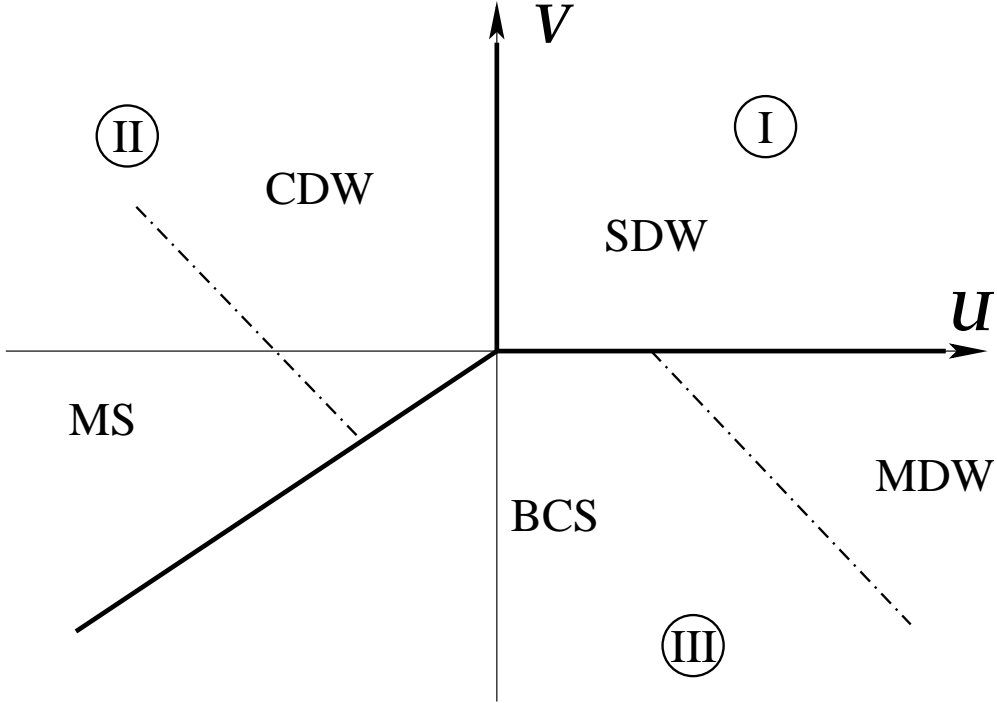}
\end{center}
\caption{
Phase diagram of model (\ref{hubbardS}) for incommensurate filling.
}
\end{figure}

\subsubsection{Critical phase}

In phase I of Fig. 2, the interaction in 
the spin sector (Eq. (\ref{spinham})) is marginally 
irrelevant when $U$ and $V$ are both 
positive so that the phase displays an extended quantum critical behavior. 
Up to a spin-velocity anisotropy, the low-energy 
properties of this phase 
are very similar to that of the repulsive SU($2N$) Hubbard chain 
with one gapless charge mode and $2N-1$ gapless spin excitations \cite{affleck,assaraf}.
All correlation functions in this phase display a power-law behavior.
The leading instabilities are those which have the 
slowest decaying correlations at long distance.
For phase I, the dominant instabilities
are the $2k_F$ CDW $\rho_{2k_F}$ 
and generalized $2k_F$ spin-density wave (SDW) ${\cal S}^{A}_{2k_F}$
order parameters which read as follows in terms 
of the Dirac fermions:
\begin{equation}
\rho_{2k_F} = L^{\dagger}_{\alpha} R_{\alpha}, \; \;
{\cal S}^{A}_{2k_F} = L^{\dagger}_{\alpha} T^{A}_{\alpha \beta} R_{\beta} .
\label{rhosdw}
\end{equation}
Using the representation (\ref{grepapp}), we obtain the leading asymptotics 
of the $2k_F$ CDW correlation function:
\begin{equation}
\langle \rho_{2k_F}^{\dagger} \left(x,\tau\right) 
\rho_{2k_F} \left(0,0\right) \rangle \sim
\left(x^2 + v_c^2 \tau^2\right)^{- K_c/2N}
\left(x^2 + v^2 \tau^2\right)
^{-\left(2N + 1 \right)/2\left(N+2\right)} 
\left(x^2 + v_0^2 \tau^2\right)^{-\left(N-1\right)/N\left(N+2\right)} ,
\label{2kfcdwcorphase1}
\end{equation}
where $v$ and $v_0$ are respectively the spin-velocity
in the Sp($2N$) and ${\mathbb{Z}}_N$ sectors.
We have obtained a similar estimate for the correlations
which involve the $2k_F$ SDW operator.
At this point, these two correlation functions have  the same
power-law decay. The logarithmic corrections will lift this 
degeneracy and we expect that the SDW operator will be the dominant
instability as in the $N=1$ case \cite{bookboso,giamarchi}.

\subsubsection{Confined phase}
Let us now 
consider the first spin-gapped phase, i.e. phase II of Fig. 1,
which occurs when $U <0$ and $V > N U/2$ in the weak-coupling limit.
In contrast to phase I, it has only one gapless mode which stems
from the criticality of the charge degrees of freedom.
The existence of a spin gap leads us to expect the emergence
of a quasi-long-range BCS phase 
with the pairing of fermions, i.e a Luther-Emery liquid phase,
as in the $F=1/2$ case \cite{bookboso,giamarchi}.
However, this is not the case for $F \ge 3/2$ due to the existence 
of the  ${\mathbb{Z}}_N$ symmetry (\ref{ZNsymmetry}) which remains unbroken
in phase II.
Indeed,
it costs a finite energy gap to excite states that break
this symmetry and the dominant instabilities must thus be
neutral under ${\mathbb{Z}}_{N}$.
In particular, there is no
dominant BCS instability in phase II since, as already stated in the
introduction, the lattice singlet-pairing
operator $P_{00,i}^{\dagger}$ (\ref{bcs})
is not invariant under the
${\mathbb{Z}}_{N}$ symmetry (\ref{ZNsymmetry}).  
Another way to see this is to use the low-energy description
of the BCS operator obtained in Appendix C (Eq. (\ref{grepbisapp})):
\begin{equation}
P^{\dagger}_{00} \sim L^{\dagger}_{\alpha} {\cal J}_{\alpha \beta} R^{\dagger}_{\beta}
 \sim
:\exp \left(i\sqrt{2 \pi/N} \Theta_c\right): \sigma_1 {\rm Tr} \; \phi^{(1)} ,
\label{bcsrep}
\end{equation}
where ${\rm Tr} \; \phi^{(1)}$ takes a non-zero expectation value
in phase II. Since the ${\mathbb{Z}}_N$ symmetry of the underlying
Ising model is not broken,
the ${\mathbb{Z}}_N$ Ising spin variable $\sigma_1$ has zero expectation value and short range correlations. Consequently, 
 the equal-time correlation function of 
the BCS instability (\ref{bcsrep})
has a short-range behavior:
$\langle P^{\dagger}_{00} \left(x\right) P_{00} \left(0\right) \rangle 
\sim e^{- x/\xi}$.
The BCS singlet pairing is completely suppressed in this phase.

In contrast,
the
dominant instabilities in phase II 
are expected to be: 
\begin{equation}
\rho_{2k_F}=
L^{\dagger}_{\alpha} R_{\alpha}, \; \;  M_0 =
\epsilon^{\alpha_1 \ldots \beta_N} R^{\dagger}_{\alpha_1} \ldots
R^{\dagger}_{\alpha_N} L^{\dagger}_{\beta_1} \ldots L^{\dagger}_{\beta_N} ,
\label{orderphase2}
\end{equation}
which are respectively the $2k_F$ CDW
and the uniform component of the lattice
SU($2N$)-singlet superconducting instability made 
of $2N$ fermions, i.e. the MS instability,
$M_i = 
c_{1,i}^{\dagger}..c_{2N,i}^{\dagger}$.  
Both order parameters (\ref{orderphase2}) are neutral 
under the ${\mathbb{Z}}_N$ symmetry (\ref{znfer})
and we notice that $M_0$ is also invariant under 
the ${\tilde {\mathbb Z}}_N$ symmetry (\ref{tildeznfer}).
The long-distance behavior of the correlation functions 
of these operators can be determined using the identifications 
(\ref{grepapp}) and (\ref{idenmsnevenapp}, \ref{idenmsnoddapp}) of Appendix C:
\begin{eqnarray}
\rho_{2k_F} &\sim& 
:\exp \left(i\sqrt{2 \pi/N} \Phi_c\right): \mu_1 {\rm Tr} \; \phi^{(1)} 
\label{rhophase2rep}
\\
M_0 &\sim& 
:\exp \left(i\sqrt{2 \pi N} \Theta_c\right):
\sum_{p=0}^{N/2} a_p \; \epsilon_{N/2 -p} \; {\rm Tr} \; \phi^{(2p)} ,
\label{msphase2rep}
\end{eqnarray}
where we have assumed, for the sake of simplicity, that 
$N$ is even for the representation of the MS instability.
The fields in the Sp($2N$) $\times$ ${\mathbb{Z}}_N$ sector,
which occur in these expressions, have non-zero expectation
values in phase II so that: 
\begin{eqnarray}
\rho_{2k_F} &\sim&
:\exp \left(i\sqrt{2 \pi/N} \Phi_c\right):
\label{rhophase2}
\\
M_0 &\sim&
:\exp \left(i\sqrt{2 \pi N} \Theta_c\right): .
\label{msphase2}
\end{eqnarray}
We then deduce that these orders have 
power-law decaying equal-time correlation functions: 
\begin{eqnarray}
\langle
\rho_{2k_F}^{\dagger} \left(x\right) \rho_{2k_F} \left(0\right) \rangle
&\sim&  x^{-K_c/N}, 
\label{domincdwhase1}
\\
\langle M^{\dagger}_0 \left(x\right) M_0 \left(0\right)
\rangle &\sim& x^{-N/K_c}.  
\label{dominstaphase1}
\end{eqnarray}
We thus see that CDW and MS instabilities compete and 
the key point of the analysis is the one which dominates.
At issue is the value of the Luttinger parameter $K_c$.
For $K_c < N$, the leading instability is
$\rho_{2k_F}$ which gives rise to a CDW
phase whereas for $K_c > N$ a MS
phase is stabilized with a lattice order parameter $M_i$.
Such a large value of $K_c$ is not guaranteed for fermionic
models with only short-range interactions as model (\ref{hubbardS}).
The full non-perturbative behavior of the Luttinger
parameter as a function of the interactions $U,V$ 
and density $n$ is beyond scope of the low-energy
approach. Its value should be determined numerically
by means of Quantum Monte Carlo (QMC) technique
or the density matrix renormalization group (DMRG) calculations.
Such analysis has been performed recently for the spin-3/2 case, i.e. 
$N=2$, and reveals that the MS phase (a quartetting
phase in that case) exists for a wide range 
of attractive contact interactions at sufficiently
small density \cite{capponiMS}.
In the general $N$ case, a strong-coupling investigation at 
small density
shows that an upper bound for $K_c$ is 
$K_{c {\rm max}} = N$ \cite{capponiMS}.
From the perturbative estimate (\ref{luttpara}), we see 
that $K_c >1$ for attractive interactions 
and increases with $|U|$ and $|V|$ 
so that we expect the emergence of a MS phase in the 
general $N$ case for sufficiently
strong attractive interactions and small density.
Such a phase is displayed on Fig. 2 and 
the doted line marks the crossover line between
the CDW and MS phases.

The MS phase with the formation of bound-states made of 
$2N$ fermions is remarkable and is a consequance of the higher-spin 
degeneracy when $F \ge 3/2$. 
The onset of this phase stems from the existence of 
the ${\mathbb{Z}}_{N}$ symmetry  which remains unbroken 
in phase II.
This discrete symmetry leads to the suppression
of the charge $2e$ BCS superconducting instability
and confines the electronic charge to
multiple of $2Ne~$ i.e. the leading superfluid instability
is a composite object made of $2N$ fermions.
In this respect, the MS phase might be viewed as
a kind of nematic Luther-Emery liquid since
SU(2) invariant spin-1/2 electronic systems with a spin 
gap and gapless charge degrees of freedom, i.e. 
Luther-Emery liquids,  exhibit a 
pairing phase \cite{seidel}. 

\subsubsection{Deconfined phase}
The properties of phase III are obtained from those of phase II
with help of the duality symmetry ${\cal D}$ (\ref{dualityfer}) on 
the fermions or $( 
\Phi_c \leftrightarrow \Theta_c, {\mathbb{Z}}_N \leftrightarrow {\tilde
  {\mathbb{Z}}}_N)$ in terms of the charge and spin degrees of freedom.  
Phase III has only one gapless charge mode and the 
${\mathbb{Z}}_{N}$ symmetry, in the spin sector, is now 
spontaneously broken.
The effective two-dimensional ${\mathbb{Z}}_{N}$ Ising model 
belongs to its low-temperature phase where 
the ${\tilde {\mathbb Z}}_N$ remains unbroken. 
We find now the emergence of a
quasi-long-range BCS pairing phase
which is described by the order parameter (\ref{bcsrep}).
Since $\sigma_1$ and ${\rm Tr} \phi^{(1)}$ acquire now
a non-zero expectation value in phase III, we have the following
leading behavior at low-energy:
\begin{equation}
P^{\dagger}_{00} 
 \sim
:\exp \left(i\sqrt{2 \pi/N} \; \Theta_c\right): ,
\label{bcsphase3}
\end{equation}
which leads to the power-law decaying equal-time correlation function:
\begin{equation}
\langle
P^{\dagger}_{00} \left(x\right) P_{00} \left(0\right) \rangle
\sim x^{- 1/NK_c} . 
\label{bcscorrphase3}
\end{equation}
The ${\mathbb{Z}}_N$ symmetry, being spontaneously broken in phase III,
does not confine anymore the electronic charge and thus 
accounts for the emergence of a BCS superfluid phase.
In the spin-3/2 case, this phase has been found numerically 
by means of QMC and DMRG calculations \cite{capponi}.
The MS instability, being neutral under ${\mathbb{Z}}_N$,
still has the power-law decay (\ref{dominstaphase1}) in phase III and is 
always subdominant with respect to the BCS operator (\ref{bcsphase3}).
In contrast, the latter singlet pairing 
instability is now competing with the operator 
${\bar \rho}^{(2N)}_{2Nk_{F}}$ which is obtained 
from the MS instability (\ref{orderphase2})
by the duality transformation (\ref{dualityfer}): 
\begin{equation}
{\bar \rho}^{(2N)}_{2Nk_{F}} = \epsilon^{\alpha_1 \ldots \alpha_N 
\beta_1 \ldots \beta_N}
{\cal J}_{\alpha_1 \gamma_1}...  {\cal J}_{\alpha_N \gamma_N}\;
R_{\gamma_1}...
R_{\gamma_N}L^{\dagger}_{\beta_1}...L^{\dagger}_{\beta_N} ,
\label{mdwphase3}
\end{equation}
which corresponds to a molecular density-wave (MDW) phase,
with wave vector $2N k_F$,
made of $2N$ fermions.
Its low-energy description can be derived from Eq. (\ref{msphase2rep})
by means of the Gaussian duality ($\Phi_c \leftrightarrow \Theta_c$)
and the KW transformation:
\begin{equation}
{\bar \rho}^{(2N)}_{2Nk_{F}}  \sim
:\exp \left(i\sqrt{2 \pi N} \Phi_c\right):
\sum_{p=0}^{N/2} a_p \; \left(-1\right)^{N/2 -p} \epsilon_{N/2 -p} 
\; {\rm Tr} \; \phi^{(2p)} ,
\label{mdwphase3rep}
\end{equation}
where we have used the KW transformation of the thermal operators
of the ${\mathbb{Z}}_N$ CFT: $\epsilon_j \rightarrow (-1)^j \epsilon_j$.
The Sp($2N$) $\times$ ${\mathbb{Z}}_N$ 
fields in Eq. (\ref{mdwphase3rep})
have non-zero expectation values in phase III so that the leading 
asymptotics of the MDW correlation reads as follows:
\begin{equation}
\langle
{\bar \rho}^{(2N) \; \dagger}_{2Nk_{F}} \left(x\right) 
{\bar \rho}^{(2N)}_{2Nk_{F}} \left(0\right) \rangle
\sim x^{- N K_c} .
\label{MDWcorrphase3}
\end{equation}
Using Eq. (\ref{bcscorrphase3}), we conclude that
the BCS instability is the dominant one of phase III 
when $K_c > 1/N$ whereas the MDW phase emerges for
$K_c < 1/N$. The crossover line between these two phases
is denoted by dotted lines in Fig. 2.

\subsubsection{Quantum phase transition}
As it has been discussed in section II, the nature of the
quantum phase transition between the two spin-gapped phases II and III
can be determined through the 
duality symmetry ${\cal D}$ (\ref{dualityfer}).  
On the
self-dual line $g_{\perp}=0$, i.e. $2V = NU$, 
the low-energy properties of the transition are captured
by the field theory (\ref{spinhamtrans}).
Using the conformal embedding approach, we observe that there is a separation of
the Sp($2N$) and ${\mathbb{Z}}_N$ degrees of freedom
in Eq. (\ref{spinhamtrans}).  Though the
Sp($2N$) sector remains gapfull when $U < 0$, the effective
${\mathbb{Z}}_N$ Ising model is at its self-dual critical point and
governs the phase transition.  
The ${\mathbb{Z}}_N$ quantum
criticality might be revealed explicitly by considering the following
ratio ${\cal R}_N$ at equal times:
\begin{equation}
{\cal R}_N \left(x\right) = \left(\langle P^{ \dagger}_{00} \left(x\right) 
 P_{00} \left(0\right)
\rangle \right)^{N^2} /\langle M^{\dagger}_0 \left(x\right) 
M_0 \left(0\right) \rangle .
\label{ratio}
\end{equation}
In phase II, this ratio admits an exponential decay since
the singlet-pairing instability is short range.
Using Eqs. (\ref{dominstaphase1}, \ref{bcscorrphase3}), 
we find that ${\cal R}_N \left(x\right) \sim {\rm cte}$ in phase III.
At the ${\mathbb{Z}}_N$ quantum critical point, 
${\cal R}_N \left(x\right)$ has a power-law behavior.
Indeed, at this point, the BCS instability (\ref{bcsrep})
simplifies as follows:
\begin{equation}
P^{ \dagger}_{00} \left(x\right)
\sim   
:\exp \left(i\sqrt{2 \pi/N} \Theta_c\right): \sigma_1 ,
\label{bcstrans}
\end{equation}
where we have averaged the Sp($2N$) degrees of freedom.
Similarly, using the identification (\ref{msphase2rep}), 
we obtain that the MS instability $M_0 \left(x\right)$ 
is still given by Eq. (\ref{msphase2}).
Therefore, we find that, at the ${\mathbb{Z}}_N$ critical point,
${\cal R}_N \left(x\right)$ has a
power-law decay with a  \emph{universal} exponent: 
\begin{equation}
{\cal R}_N \left(x\right) \sim x^{-2 N(N - 1)/(N + 2)},
\label{ratiotrans}
\end{equation}
i.e.  an exponent $1$ and $12/5$ 
respectively for the Ising ($N=2$) and three-state Potts ($N=3$) cases. 
The long-distance behavior of the 
function ${\cal R}_2 \left(x\right)$  close to the transition
has been determined recently numerically by means of 
large scale DMRG calculations
and the emergence of the ${\mathbb{Z}}_2$ quantum criticality 
has been revealed \cite{capponi}.
However,  the phase
transition may be non-universal for larger $N$ 
since we have neglected irrelevant perturbations in the weak-coupling
limit $|U,V| \ll t$ that could become relevant at the quantum
phase transition.
Some insights might be gained from the symmetries of the problem.
The relevant operators should belong to the ${\mathbb{Z}}_N$ sector,
be neutral under the ${\mathbb{Z}}_N \times {\tilde {\mathbb{Z}}}_N$
symmetry and invariant under the KW transformation.
The natural candidates are the thermal 
operators $\epsilon_j$ ($j=1,\ldots, N/2$)
of the ${\mathbb{Z}}_N$ CFT with scaling dimension $2j(j+1)/(N+2)$
which transforms as $\epsilon_j \rightarrow (-1)^j \epsilon_j$
under the KW duality symmetry \cite{para}.
We thus deduce that, in the $N=2,3$ cases, the quantum phase
transition between phases II and III is universal and belongs
to the Ising and three-state Potts universality classes respectively.
For $N \ge 4$, a strongly relevant
perturbation $\epsilon_2$ with scaling dimension $12/(N+2)$
is generated in the ${\mathbb{Z}}_N$ sector.
The resulting field theory which captures the quantum phase transition 
for $F \ge 7/2$ becomes: 
\begin{equation}
{\cal S}_{\rm eff} = {\cal S}_{{\mathbb{Z}}_N}  + \lambda 
\int d^2 x \; \epsilon_2 \left(x\right) , 
\label{efftrans}
\end{equation}
where ${\cal S}_{{\mathbb{Z}}_N}$ stands for the action
of the ${\mathbb{Z}}_N$ CFT.
Model (\ref{efftrans}) turns out to be an integrable deformation
of the ${\mathbb{Z}}_N$ CFT \cite{fateev}.
The nature of the phase transition depends on 
the sign of the coupling
constant $\lambda$ \cite{fateev}.
For $\lambda <0$, the field theories (\ref{efftrans}) are
massive and the phase transition is of first-order type.
For $\lambda  > 0$ 
it is known that model (\ref{efftrans})
has a massless RG flow
onto a Kosterlitz-Thouless (KT) U(1) gapless phase 
with central charge $c=1$. 

Finally,  we end this subsection by 
briefly considering the other phase transitions 
that occur in Fig. 2.
The quantum phase transition between phases I and III
is of KT type and its exact position is $V=0$ and $U > 0$, as it
will be seen in section III B 4,
which corresponds to the repulsive SU($2N$) Hubbard model.
The transition between phases I and II is also expected
to be of the  KT type but its actual position is 
not clear. Within the weak-coupling approach of section II, 
its location is $U=0$ and $V>0$ but higher-order corrections
might bend this line.

\subsection{Commensurate filling: one-atom per site}

We now discuss the nature of the phase diagram of
model (\ref{hubbardS}) for a 
commensurate $1/2N$ filling ($k_F = \pi/(2N a_0)$) i.e. one atom
per site. 
In this case, there is still a spin-charge separation 
(\ref{spinchargesepar}) and
the main modification
occurs in the charge sector with the existence of an umklapp
process which appears at higher order of the perturbation
theory,
and that transfers $2N$ fermions from one Fermi 
point to the other, i.e. $L^\dagger_1R_1L^\dagger_2R_2\dots L^\dagger_{2N}R_{2N}+\mbox{H.c.}$.
The low-energy Hamiltonian, which describes the charge
degrees of freedom, becomes now a sine-Gordon model:
\begin{equation}
{\cal H}_c = \frac{v_c}{2} \left[\frac{1}{K_c}
\left(\partial_x \Phi_c \right)^2 + K_c
\left(\partial_x \Theta_c \right)^2 \right]
- g_c \cos \left( \sqrt{8\pi N} \; \Phi_c \right) .
\label{sinegordon}
\end{equation}
Such umklapp operator can also be determined
from a symmetry analysis.
Indeed,
for the commensurate filling $k_F = \pi/(2N a_0)$, using the representation
(\ref{contfer}), the lattice one-step
translation symmetry ${\cal T}_{a_0}$ reads as follows
on the Dirac fermions:
\begin{equation}
L_{\alpha} \rightarrow e^{- i \pi/2N} L_{\alpha}, \; \;
R_{\alpha} \rightarrow e^{i \pi/2N} R_{\alpha} ,
\label{trans}
\end{equation}
so that we deduce $\Phi_c \rightarrow \Phi_c + \sqrt{\pi/2N}$ 
under ${\cal T}_{a_0}$ from Eq. (\ref{fernonabelboso}).
The cosine term of Eq. (\ref{sinegordon}) is thus 
the one with the smallest scaling dimension and
compatible with the translation symmetry.
Since its scaling dimension
is $\Delta_u = 2 N K_c$, we
deduce that  a charge gap is 
opened when $K_c < 1/N$.
A Mott transition occurs between a Luttinger phase
and  an insulating phase (see Fig. 3).
In the latter phase, the charge bosonic field $\Phi_c$ 
is pinned in one of the minima of 
the sine-Gordon model (\ref{sinegordon}):
\begin{eqnarray}
\langle \Phi_c \rangle &=& \sqrt{\frac{\pi}{2N}} \; m , \; \; g_c > 0
\nonumber \\
\langle \Phi_c \rangle &=& \sqrt{\frac{\pi}{2N}} 
\; \left( m + \frac{1}{2} \right) ,
\; \; g_c < 0 ,
\label{pinning}
\end{eqnarray}
$m$ being integer.
Three different Mott-insulating phases 
can then be defined depending on the spin degrees of freedom 
and the status of the ${\mathbb{Z}}_N$ discrete symmetry (see Fig. 3).

\begin{figure}
\begin{center}
\includegraphics[width=0.8\linewidth]{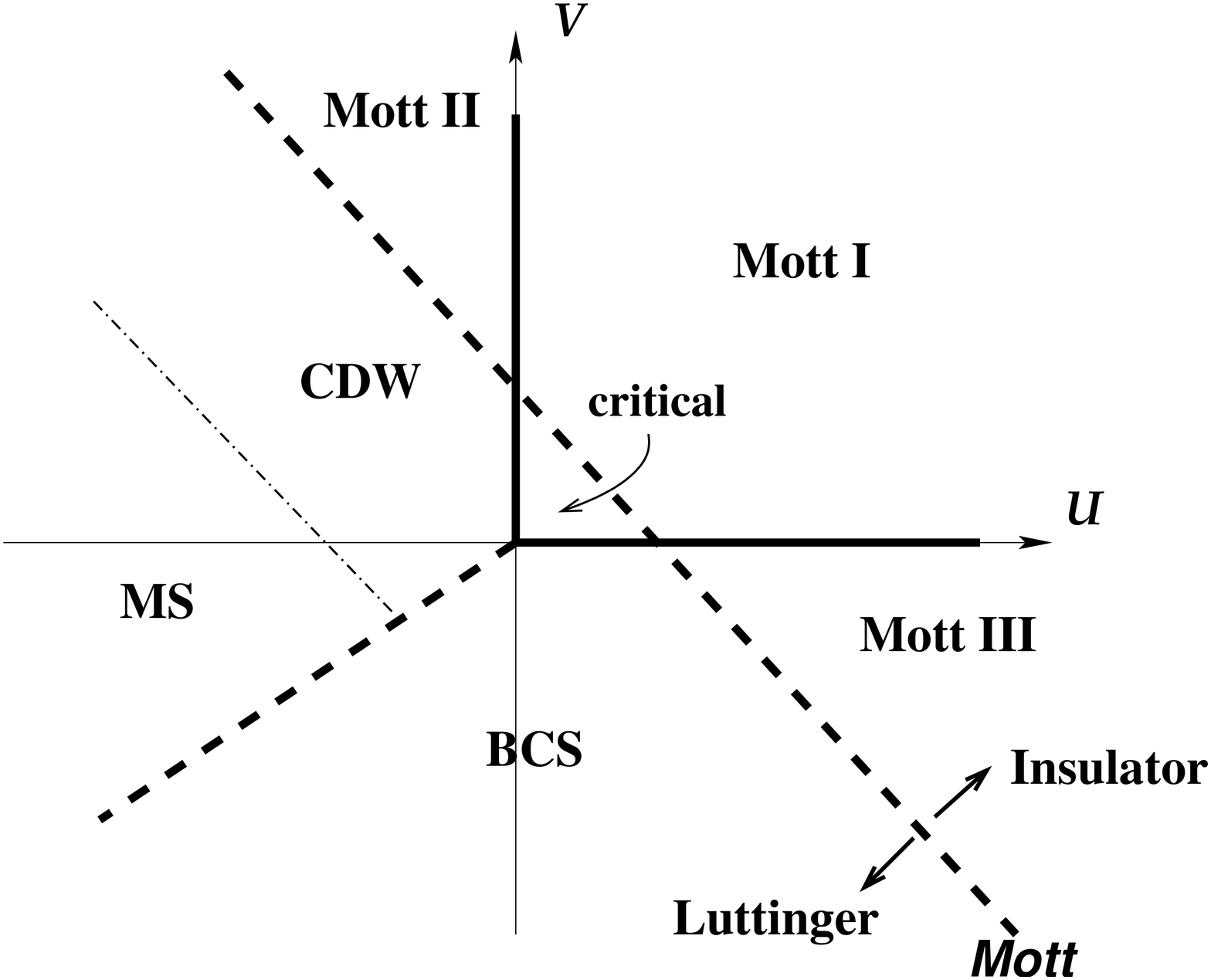}
\end{center}
\caption{
Phase diagram of model (\ref{hubbardS}) for a filling of one atom per site.
}
\end{figure}

\subsubsection{Mott I phase}
\label{mott1}
In the first Mott phase, 
the spin degrees of freedom remain gapless.
This phase is qualitatively similar to the insulating phase of
the repulsive SU($2N$) Hubbard chain at $1/(2N)$ filling with
$2N -1$ gapless bosonic spin modes \cite{affleck,assaraf}. 
In particular, it includes (when taking the limit of large repulsive $U$) the SU(2N) generalization of the 
Heisenberg model:
\begin{equation}
{\cal H}_{\rm } = J \sum_i S^A_i S^A_{i+1},
\label{sutherlandmodel}
\end{equation}
where $S^A_i$ are the SU($2N$) 
spin operators on site $i$:
$S^A_i = c^{\dagger}_{\alpha,i} 
T^A_{\alpha \beta} c_{\beta,i}$.
This model is integrable by means of 
the Bethe ansatz approach \cite{sutherland}
and its critical properties are 
captured by the SU($2N$)$_1$ WZNW model \cite{affleck}.
The leading asymptotics of the spin-spin correlation functions in this 
phase can be easily obtained from Eq. (\ref{2kfcdwcorphase1}) 
by gapping the charge 
degrees of freedom:
\begin{equation}
\langle S^A \left(x,\tau\right)
S^B \left(0,0\right) \rangle \sim
\frac{\delta^{AB} \cos\left( 2k_F x \right)}{
\left(x^2 + v^2 \tau^2\right)
^{\left(2N + 1 \right)/2\left(N+2\right)}
\left(x^2 + v_0^2 \tau^2\right)^{\left(N-1\right)/N\left(N+2\right)}} .
\label{spincorr}
\end{equation}

\subsubsection{Mott II phase}
The remaining Mott insulating phases of Fig. 3 are fully gapped.
In region II, we have seen in section 
III that the ${\mathbb{Z}}_N$ (respectively 
${\tilde {\mathbb{Z}}}_N$) symmetry in
the spin sector is unbroken (respectively spontaneously broken).
We expect that the lattice translation symmetry ${\cal T}_{a_0}$ is
spontaneously broken in the Mott II phase.
This can be explicitly shown by considering the lattice 
$2N$-merization operator which is defined by:
\begin{equation}
\Delta_{2k_F}   =  e^{- i \pi j/N} S^A_{j} S^A_{j+1} .
\label{2Nmer}
\end{equation}
In the continuum limit, this order parameter expresses directly in terms 
of the SU($2N$)$_1$ primary field:
$\Delta_{2k_F}   \sim {\rm Tr} g \sim \mu_1 {\rm Tr} \phi^{(1)}$.
Since the ${\tilde {\mathbb{Z}}}_N$ symmetry is broken, 
i.e. $\langle \mu_1 \rangle \ne 0$, 
${\rm Tr} g$ acquires a non-zero expectation value 
in this phase:
$\langle \Delta_{2k_F} \rangle \ne 0$.
We deduce, from the definition (\ref{2Nmer}), 
that the translation symmetry is spontaneously broken 
and the Mott II phase 
displays a $2N$-merization with a $2N$ ground-state degeneracy.
On top of this order,
we also expect the formation of a long-range ordering of a 2k$_{F}$ CDW 
with order parameter $\rho =  \cos\left(\pi j/N\right) \;
c^{\dagger}_{\alpha,j} c_{\alpha,j}$.
Using Eqs. (\ref{contfer},\ref{rhophase2}), we get 
the following identification in phase II: 
\begin{equation}
\rho \simeq 
L^{\dagger}_{\alpha} R_{\alpha} + 
R^{\dagger}_{\alpha} L_{\alpha} \sim \;
:\cos \left(\sqrt{2 \pi/N} \; \Phi_c\right): ,
\label{mott2elcdw}
\end{equation}
which has a non-zero expectation value due to the pinning of
the charge bosonic field (\ref{pinning}).
The 2k$_{F}$ CDW coexists with the $2N$-merization operator (\ref{2Nmer})
in the Mott II phase.

\subsubsection{Mott III phase}
The Mott transition line $K_c = 1/N$ coincides with 
the cross-over line between BCS and MDW phases in Fig. 2. 
For a filling of one atom per site, the BCS phase is still present while
the MDW phase becomes 
a fully gapped Mott insulating phase.
In this Mott phase, the ${\mathbb{Z}}_N$ symmetry
is now spontaneously broken while 
the ${\tilde {\mathbb{Z}}}_N$ one remains unbroken.
In the continuum limit, an order parameter of this phase 
is provided by the $2N k_F$ MDW operator (\ref{mdwphase3}) 
or the $2N k_F$ CDW: 
$\rho_{2Nk_F} = \epsilon^{\alpha_1 \ldots \alpha_N
\beta_1 \ldots \beta_N} L^{\dagger}_{\alpha_1}
\ldots L^{\dagger}_{\alpha_N} 
R_{\beta_1} \ldots R_{\beta_N}$.
In particular, using the character decomposition (\ref{chardecoms}) 
of Appendix C, 
we get:
\begin{equation}
\rho_{2Nk_F} \sim :\exp \left(i\sqrt{2 \pi N} \Phi_c\right):
\sum_{p=0}^{N/2} a_p \; \epsilon_{N/2 -p} \; {\rm Tr} \; \phi^{(2p)} .
\label{2Nkfcdwrep}
\end{equation}
After averaging out the Sp($2N$)$_1$ $\times$
${\mathbb{Z}}_N$ fields, $2N k_F$ MDW (\ref{mdwphase3rep}) and 
$2N k_F$ CDW  order parameters have
the same leading behavior which is given by:
\begin{equation}
\rho_{2Nk_{F}}  \sim
:\exp \left(i\sqrt{2 \pi N} \; \Phi_c\right): .
\label{mdwpmott}
\end{equation}
Using the position of the 
pinning of the charge bosonic field (\ref{pinning}),
we deduce that 
$\langle {\rm Re} \; \rho_{2Nk_{F}} \rangle \ne 0$
(respectively 
$\langle {\rm Im} \; \rho_{2Nk_{F}} \rangle \ne 0$) 
when $g_c > 0$ (respectively $g_c  < 0$). 
The electronic properties of this phase thus depends on
the sign of the coupling
constant $g_c$ of the umklapp operator in Eq. (\ref{sinegordon}).
This sign might be fixed using higher-order perturbation 
theory as in Ref. \cite{orignac} or numerically 
by looking at the different lattice order parameters of the 
problem. In this respect, for $g_c > 0$, 
we have the long-range ordering of a $q=2Nk_F=\pi/a_0$ 
CDW: $\langle \rho_{\pi} \rangle = \langle \sum_{j,\alpha} (-1)^j 
c^{\dagger}_{j, \alpha} c_{j, \alpha} \rangle \ne 0$ while 
for $g_c < 0$ a spin-Peierls (bond) ordering is formed:
$\langle {\cal O}_{\rm SP} \rangle = 
\langle \sum_{j,\alpha} (-1)^j
c^{\dagger}_{j+1, \alpha} c_{j, \alpha} + \mbox{H.c.} \rangle \ne 0$.
In any case, the Mott III phase is two-fold degenerate
for all $N$ and the translation symmetry is broken.
This Mott phase is quite unusual since
there is no one-particle charge density long-range fluctuation due to the
spontaneously breaking  of the ${\mathbb{Z}}_N$ 
symmetry. 
In the $N=2$ case, at quarter-filling, it has been 
found numerically by means of the DMRG approach that 
a Mott transition occurs with the formation of a 
$4 k_F = \pi/a_0$ bond ordering insulating 
phase \cite{capponi}. We thus expect, at least for 
$N=2$, that the coupling constant 
of the umklapp process is negative in phase III: $g_c <0$.
Finally,  we observe that the
Mott III phase is the only fully gapped phase directly 
connected to the BCS phase (see Fig. 3).

\subsubsection{Strong coupling expansion}
At large positive $U$, charge degrees of freedom 
are completely frozen by a large Mott gap, and it is possible to perform a strong coupling expansion
that leads to an effective model for the 
low-energy spin degrees of freedom. Indeed, when $t/U=0$ in Eq. (\ref{hubbardS}), 
the ground states with energy 0 (the chemical potential 
is set to $\mu= U/2$), which 
consist of states with all sites 
filled by exactly one 
particle (`singly-occupied states'), are well separated 
in energy from all other states by a 
large gap $U$. One should note that the gap to those states 
that are singly occupied on all sites except 
on two sites, one being empty and the other one being 
doubly occupied by a SU(2) singlet $P^\dagger_{00}\left|0\right\rangle$, 
is $U+V$. Therefore the following picture holds 
only when $t\ll U,U+V$ to avoid formation 
of BCS pairs that occurs at large negative $V$. Nevertheless, we shall see that 
the BCS phase is captured by this approach, since it extends up to $V=0^{-}$.

The huge degeneracy of the ground-state manifold consisting of 
singly occupied states is lifted by the 
hopping term ${\cal H}_0=-t\sum_{i,\alpha}c^\dagger_{\alpha,i}c_{\alpha,i} + \mbox{H.c.}$ 
as soon as $t/U$ departs from 0. 
To second order (the first non-trivial order) 
in perturbation theory, the effective Hamiltonian acting on singly occupied states reads:
\begin{equation}
{\cal H}_{\mbox{eff}}=\mathcal P
{\cal H}_0(1-\mathcal P)\frac{1}{{\cal H}_{U,V}}(1-\mathcal P) {\cal H}_0P,
\label{HeffstrongU}
\end{equation}
where $\mathcal P$ is the projector onto singly 
occupied states and ${\cal H}_{U,V}= {\cal H} - {\cal H}_0$ is 
the quartic part of the Hamiltonian (\ref{hubbardS}).
The effective Hamiltonian can then be expressed in terms of SU($2N$) spin 
operators defined on each site, $S^A_i=c^\dagger_{\alpha,i}T^A_{\alpha\beta}c_{\beta,i}$, 
out of which we can single out the $S^a_i$ ($a=1, \ldots, N(2N+1)$) that constitute 
a basis of the Sp($2N$) spin operators. One finds:
\begin{equation}
{\cal H}_{eff} =
G_0 \sum_{i} \sum_{A=1}^{4N^2-1}S^A_i S^A_{i+1} 
+ G_1 \sum_{i} \sum_{a=1}^{N(2N+1)}S^a_i S^a_{i+1} ,
\label{effhamstrong}
\end{equation}
with the couplings 
$G_0=\frac{t^2}{N}\left(\frac{2N-1}{U} + \frac{1}{U+V}\right)$ 
and $G_1=\frac{t^2}{N}\left(\frac{1}{U}-\frac{1}{U+V}\right)$.

When $V=0$, $G_1=0$ and the model reduces to the SU($2N$) 
antiferromagnet  in the fundamental representation. As recalled 
in section \ref{mott1}, this integrable model is described at low energy 
by the SU$(2N)_1$ WZNW model perturbed by a marginally 
irrelevant current-current interaction. At small $V/U$, i.e. in the vicinity of 
the SU($2N$) invariant point, we can thus obtain a continuum limit 
of Hamiltonian (\ref{effhamstrong}) in the form of a 
perturbed SU$(2N)_1$ WZNW model. The calculation relies on the continuous 
representation of the spin operators, 
$S^A_j\simeq S^A(x=ja_0)=I^A_L+I^A_R
+\sum_{m=1}^{2N-1}\alpha_m e^{2imk_Fx}{\cal S}^A_{2mk_F}$, 
where the numbers $\alpha_m$ are non-universal amplitudes. 

This procedure yields a low-energy Hamiltonian density 
exactly of the form (\ref{spinham}), 
with couplings $g_\parallel(U,V)$ and $g_\perp(U,V)$. 
The actual value of $g_\parallel$ and $g_\perp$ are not 
known (they depend on non-universal amplitudes), 
but it is sufficient to know that $g_\parallel(U,0)=g_\perp(U,0)<0$ 
at the SU($2N$) invariant point, and 
that $g_\parallel-g_\perp\propto G_1\propto V$ 
at small $V/U$. The RG analysis of section \ref{RGanalysis} 
thus leads to the conclusion that the spin sector 
undergoes a transition at $V=0$, between a 
quantum critical phase ($V>0$), that coincides with 
the Mott phase I, and a spin gapped phase ($V<0$) with 
unbroken ${\tilde {\mathbb Z}}_N$ symmetry, which is 
Mott phase III. The strong coupling expansion thus 
establishes that the exact position of the transition 
line between these two Mott phases is $U>0,V=0$.

\section{Concluding remarks}

In summary, we have studied the zero-temperature phase diagram
of one-dimensional 
spin-$F=N-1/2$ fermionic cold atoms with 
contact interactions.
In the low-energy limit, a CFT approach has been developed
to deduce the main physical properties of the model
by exploiting the presence of an extended Sp($2N$) symmetry. 
In the $F=3/2$ case, this Sp($4$) $\sim$ SO($5$) symmetry
is exact on the lattice while for larger $F$ its existence
stems from a fine-tuning of the coupling constants i.e. 
the scattering lengths of the atoms.
	
The phase diagram of the model is very rich for 
incommensurate filling with the competition 
between instabilities of very different nature.
In particular, two different superfluid phases 
are found for attractive 
interactions: a Sp($2N$)-singlet BCS pairing phase 
and a MS phase.
The latter phase is an SU($2N$) singlet 
which is formed from bound states of $2N$ fermions.
For instance, for $N=2$, i.e. $F=3/2$, it is the analogous 
of the $\alpha$ particle in nuclear physics.
At the heart of this competition is a
${\mathbb{Z}}_N$  discrete symmetry which
is the coset between the center of the  SU($2N$) group
and the center of the  Sp($2N$) one. 
Since Sp($2N$) and SU(2) share the same
center group, this ${\mathbb{Z}}_N$ symmetry 
is not an artifact of the extended 
Sp($2N$) symmetry for $F \ge 7/2$ and
is present for general SU$(2)$-invariant multicomponent
Fermi gas. The ${\mathbb{Z}}_N$ symmetry plays a
crucial role and provides a new important 
ingredient which is not present for $F=1/2$.

In the low-energy approach,
this discrete symmetry corresponds to 
the symmetry of the low-temperature phase
of an effective two-dimensional ${\mathbb{Z}}_N$
generalized Ising model. 
If the ${\mathbb{Z}}_N$ symmetry is not spontaneously broken,
we find the suppression of the singlet-pairing BCS phase
with the confinement of the electronic charge $e$ to multiple
of $2Ne$.
In this case, the nature of the dominant instability of this phase
depends on the non-universal Luttinger parameter $K_c$. 
If $K_c < N$, the leading instability is a $2k_F$
CDW while a quasi-long-range  MS phase emerges for $K_c > N$ 
and should be the generic phase for attractive interactions
at sufficiently small density.
Due to its confinement properties and 
the existence of a spin gap, this MS phase can also be 
viewed as nematic Luther-Emery liquid.
When the ${\mathbb{Z}}_N$ symmetry is spontaneously broken, 
a quasi-long-range BCS phase is stabilized 
when $K_c > 1/N$ whereas a MDW phase appears for 
$K_c < 1/N$. 
The quantum phase transition between the confined and
BCS phases is described by the ${\mathbb{Z}}_N$
parafermionic CFT perturbed by the second thermal operator.
For $F=3/2$ and $5/2$, the transition is universal 
and belongs respectively to the Ising and three-state Potts
universality classes. For higher spins $F \ge 7/2$, 
the transition is non-universal and is either 
of first-order or in the U(1) KT universality class.

For commensurate filling of one atom per site, in sharp
contrast to the $F=1/2$ case,
a Mott transition is predicted to 
occur for $F \ge 3/2$ as the result of the spin
degeneracy. Three different Mott-insulating phases 
are expected which depend on the status 
of the ${\mathbb{Z}}_N$ symmetry.
A Mott phase with $2N-1$ gapless spin
modes is first found which is qualitatively 
similar to the insulating phase of the SU($2N$)
Hubbard chain. The two others have a spin gap
and display spin-Peierls (bond) ordering
with  a $2N$ or a two-fold ground-state 
degeneracy. 

Regarding perspectives, it will be interesting 
to investigate numerically the phase diagram
for $F=5/2$ spins by means of large-scale
DMRG or QMC calculations \cite{capponipreprint}.
In particular, the three-state Potts universal behavior of the 
quantum phase transition might be confirmed numerically
as well as the study of the Mott transition and the 
different Mott-insulating phases for 
one atom per site.
From the theoretical point of view,
another important question is the phase diagram for incommensurate
filling of the SU(2) model (\ref{hubbardSgen}) for $N \ge 3$ without any  
Sp($2N$) fine-tuning.
Finally, we hope that the exotic MS phase
for general  $F=N -1/2$ spins, discussed in this paper,
will be observed in future experiments in ultracold
spinor fermionic atoms.

\begin{acknowledgements}
The authors are very grateful to
S. Capponi, G. Roux, and S. R. White for related collaborations
on this project.
One of us (P.L.) would like also to thank V. A. Fateev and R. Santachiara
for useful discussions.  
\end{acknowledgements}

\appendix
\section{Group theory conventions}
The aim of this Appendix is to define the algebraic conventions
used in the low-energy approach
as well as the RG calculations which are presented in Appendix B.

Let us start with the SU($2N$) group.
The generators of the Lie algebra of SU($2N$) 
in the fundamental representation ($2N \times 2N$
matrices)
are denoted by
$T^A$ $(A=1,\ldots, 4N^2 -1)$
with the normalization:
$\mbox{Tr}(T^A T^B) = \delta^{AB}/2$.
They satisfy the commutation relation:
\begin{equation}
\left[T^A,  T^B \right] = i f^{ABC}  T^C,
\label{su2nrel}
\end{equation}
$f^{ABC}$ being the structure constants of SU($2N$).
An useful group identity for the derivation of the continuum limit of the 
lattice model (\ref{hubbardS}) is:
\begin{equation}
\sum_A T^A_{\alpha \beta} T^A_{\gamma \rho} = 
\frac{1}{2} \left(\delta_{\alpha \rho} \delta_{\beta \gamma}
- \frac{1}{2N}\; \delta_{\alpha \beta} \delta_{\gamma \rho} \right)  .
\label{sumsu2ngen}
\end{equation}

The SU($2N$) generators can be divided in generators 
of the subalgebra Sp($2N$), $T^a$ $(a=1,\ldots, N(2N +1))$, 
and generators of the complement of Sp($2N$) in SU($2N$): 
$T^i$ $(i=1,\ldots, 2N^2 - N - 1)$. 
They are normalized as:
$\mbox{Tr}(T^a T^b) = \delta^{ab}/2$, $\mbox{Tr}(T^i T^j) = \delta^{ij}/2$.
The various structure constants are then given by:
\begin{equation}
\left[T^a,  T^b \right] = i f^{abc}  T^c,
\left[T^i,  T^j \right] = i f^{ija}  T^a,
\left[T^a,  T^i \right] = i f^{aij}  T^j ,
\label{su2ngrad}
\end{equation}
which defines a ${\mathbb{Z}}_2$ graduation of the SU($2N$) Lie algebra. 
These coefficients can be determined by expressing the generators $T^a, T^i$
in terms of a direct product between SU($N$) and SU(2) generators.
The SU($N$) generators admit a simple $N \times N$ 
matrix representation which falls
into three categories:
\begin{itemize}
\item {\rm Symmetric part:}
\begin{equation}
\left(M^{\rm{(1)}}_{ij}\right)_{\alpha\beta}
=\frac{1}{2}(\delta_{i\alpha}\delta_{j\beta}
+ \delta_{i\beta}\delta_{j\alpha})
\qquad (1\leq i < j \leq N)
\label{eqn:SUN-gen-1}
\end{equation}
\item {\rm Antisymmetric part:}
\begin{equation}
\left(M^{\rm{(2)}}_{ij}\right)_{\alpha\beta}
=-\frac{i}{2}(\delta_{i\alpha}\delta_{j\beta}
-\delta_{i\beta}\delta_{j\alpha})
\qquad (1\leq i < j \leq N)
\label{eqn:SUN-gen-2}
\end{equation}
\item {\rm Cartan generators:}
\begin{equation}
 \left(M^{\rm{D}}_{m}\right)_{\alpha\beta}
=\frac{1}{\sqrt{2m(m+1)}}
\left(
\sum_{k=1}^{m}\delta_{\alpha k}\delta_{\beta k}
-m\, \delta_{\alpha,m+1}\delta_{\beta,m+1}
\right),  (m=1,\ldots,N-1) .
\label{eqn:SUN-gen-3}
\end{equation}
\end{itemize}

The Sp(2$N$) generators $T^a$ $(a=1,\ldots, N(2N +1))$ are then 
given by the following set:
\begin{equation}
T^a = \left\{ \frac{1}{\sqrt{2}} \; M^{\rm{(1)}}_{ij} \otimes {\vec \sigma}, 
\; \frac{1}{\sqrt{2}} \; M^{\rm{(2)}}_{ij} \otimes I_2, \; 
\frac{1}{\sqrt{2}} \; M^{\rm{D}}_{m} \otimes {\vec \sigma}, 
\; \frac{1}{2 \sqrt{N}} I_N \otimes {\vec \sigma} 
\right\}   , 
\label{sp2ngen}
\end{equation}
${\vec \sigma}$ being the Pauli matrices.
The metric ${\cal J}_{\alpha \beta}$, which defines the Sp(2$N$) group,
takes a simple form in this scheme:
${\cal J} = I_N \otimes (-i \sigma_2)$, and 
is thus the generalization of the antisymmetric tensor $\epsilon_{\alpha \beta}$ for
higher spin $F > 1/2$.
Using the representation (\ref{sp2ngen}), 
we also obtain  the following identity which is useful for the derivation
of the low-energy approach of section II: 
\begin{equation}
\sum_a T^a_{\alpha \beta} T^a_{\gamma \rho} =
\frac{1}{4} \left(\delta_{\alpha \rho} \delta_{\beta \gamma}
- {\cal J}_{\alpha \gamma} {\cal J}_{\beta \rho} \right) .
\label{sumsp2ngen}
\end{equation}
Finally, the remaining generators $T^i$ $(i=1,\ldots, 2N^2 - N - 1)$ read as follows:
\begin{equation}
T^i = \left\{ \frac{1}{\sqrt{2}} \; M^{\rm{(2)}}_{ij} \otimes {\vec \sigma},
\; \frac{1}{\sqrt{2}} \; M^{\rm{(1)}}_{ij} \otimes I_2, \;
\frac{1}{\sqrt{2}} \; M^{\rm{D}}_{m} \otimes  I_2
\right\}   .
\label{complementgen}
\end{equation}

\section{Two-loop RG results}
In this Appendix, we present 
the two-loop RG equations of the current-current model (\ref{spinham}): 

\begin{eqnarray}
{\dot g}_{\parallel} &=&
\left(1 - \frac{g_{\parallel}}{4\pi} \right)
\left[\left(N + 1\right) \frac{g_{\parallel}^2}{4\pi}
+ \left(N - 1\right) \frac{g_{\perp}^2}{4\pi}\right] \nonumber \\
{\dot g}_{\perp} &=& N \frac{g_{\perp}g_\parallel}{2\pi}  
- \frac{Ng_\perp}{(4\pi)^2}\left( g_\parallel^2+g_\perp^2\right),
\label{RGeqs}
\end{eqnarray}
where ${\dot g}_{\perp,\parallel} = 
\partial g_{\perp,\parallel}/\partial t$,
$t$ being the RG time and we have neglected the spin-velocity anisotropy.
To establish this result, we use the notations of 
Ref. \cite{leclair} (we modify normalizations for our purpose) and define the operators:
\begin{equation}
\mathcal O_\parallel(z,\bar z)=I^a_{\parallel R}(\bar z)I^a_{\parallel L}(z),\quad
\mathcal O_\perp(z,\bar z)=I^i_{\perp R}(\bar z)I^i_{\perp L}(z),
\end{equation}
in terms of which the interacting part of the action reads
\begin{equation}
S_I=\int d^2x \;g_\alpha \mathcal O_\alpha.
\end{equation}
Structure constants $C,D,\tilde C$ are defined for the algebra 
of the local operators $\mathcal O_\alpha$ and associated pseudo stress-energy 
tensors $\mathcal T_\parallel(z)=:I^a_{\parallel L}I^a_{\parallel L}:(z)$ 
and $\mathcal T_\perp(z)=:I^i_{\perp L}I^i_{\perp L}:(z)$, according to the OPE's:
\begin{eqnarray}
\mathcal O_\alpha (z,\bar z)\mathcal O_\beta (0,0) &\sim &
\frac{1}{8\pi^2\,|z|^2}\;C_{\alpha\beta}^\gamma\mathcal O_\gamma (0,0),\\
\mathcal T_\alpha (z)\mathcal O_\beta (0,0) &\sim &
\frac{1}{8\pi^2\,z^2}\;\left(2kD_{\alpha\beta}^\gamma + \tilde C_{\alpha\beta}^\gamma\right)\mathcal O_\gamma (0,0).
\end{eqnarray}
Here, to disentangle the structure constants $D$ and $\tilde C$, 
we need to vary fictitiously the level $k$ of the 
Kac-Moody algebras ($k=1$ in our case), thus considering 
the more general problem of perturbing a SU$(2N)_k$ model by 
marginal current-current interactions built on the currents 
of the Sp$(2N)_k$ that are embedded in it. We still write 
$I^A_{L,R}$ for the SU$(2N)_k$ currents. The only formal 
difference appears in the current-current OPE's:
\begin{equation}
I_L^A(z)I_L^B(0)\sim \frac{k\,\delta^{AB}}{8\pi^2\,z^2}+\frac{if^{ABC}}{2\pi\, z}\;I_L^C(0),
\end{equation}
with a similar expression for right currents. 

The two-loop $\beta$-function takes a 
compact form in terms of the structure constants \cite{leclair}:
\begin{equation}
\dot{g}_\alpha = -\frac{1}{8\pi}\, C_\alpha^{\beta\gamma} \,g_\beta g_\gamma
-\frac{k}{32\pi^2}\, \tilde C_\alpha^{\beta\gamma}D_{\beta}^{\mu\nu}\,g_\gamma g_\mu g_\nu.
\label{2bouclegen}
\end{equation}
The structure constants are 
calculated to be (we give only the non vanishing entries):
\begin{eqnarray}
\begin{array}{c}
C_\parallel^{\parallel,\parallel}=-2(N+1),\quad 
C_\parallel^{\perp,\perp}=-2(N-1),\quad 
C_\perp^{\parallel,\perp}=C_\perp^{\perp,\parallel}=-2N\\
\tilde C_\parallel^{\parallel,\parallel}=2(N+1) ,\quad 
\tilde C_\perp^{\perp,\perp}=2N ,\quad 
\tilde C_\parallel^{\parallel,\perp}=2(N-1) ,\quad
\tilde C_\perp^{\parallel,\perp}=2N ,\quad \\
D_\parallel^{\parallel,\parallel}=D_\perp^{\perp,\perp}=1 ,
\end{array}
\end{eqnarray}
so that Eq. (\ref{2bouclegen}) gives the two-loop result (\ref{RGeqs}).

\section{${\mathbb{Z}}_N$ parafermionic CFT}

In this Appendix, we discuss the ${\mathbb{Z}}_N$ parafermionic CFT 
and present the technical details
of the conformal embedding approach of 
section II which are used in section III
for the determination of the phase diagram
of model (\ref{hubbardS}).

The parafermionic CFT describes the critical properties of 
two-dimensional ${\mathbb{Z}}_N$ generalization of the Ising model.
The lattice spin $\sigma_r$ of such a model takes values:
$e^{i 2 \pi m /N }, m=0, \ldots, N-1$ and the corresponding Hamiltonian 
is ${\mathbb{Z}}_N$ invariant.
Using the KW duality symmetry, the spins $\sigma_r$ are replaced
by the dual spins $\mu_{\tilde r}$ which belong 
to the dual lattice and they define
a ${\tilde {\mathbb{Z}}}_N$ symmetry.
Two-dimensional ${\mathbb{Z}}_N$ Ising models display
two gapped phases, i.e. the ordered (respectively disordered) 
phase where the ${\mathbb{Z}}_N$  (respectively ${\tilde {\mathbb{Z}}}_N$)
symmetry is spontaneously broken, as well as
a quasi-long-range KT phase for prime $N \ge 5$.
The critical properties, along the self-dual manifold, 
are captured by the ${\mathbb{Z}}_N \times {\tilde {\mathbb{Z}}}_N$
parafermionic CFT with central charge $c= 2 (N-1)/(N+2)$ \cite{para}.
In the scaling limit, the conformal fields $\sigma_k$ and 
$\mu_k$ $(\sigma_k^{\dagger} = \sigma_{N-k}, \mu_k^{\dagger} = \mu_{N-k})$
with scaling dimensions $d_k = k(N-k)/N(N+2)$
$(k=0, \ldots, N -1)$ 
describe the long-distance correlations of $\sigma^{k}_r$
and $\mu^{k}_{\tilde r}$.
In particular, they carry respectively  
a $(k,0)$ and $(0,k)$ charge under 
the ${\mathbb{Z}}_N \times {\tilde {\mathbb{Z}}}_N$ symmetry:
\begin{eqnarray}
\sigma_k &\rightarrow& e^{i 2 \pi m k/N} 
\sigma_k \; \; {\rm under} \; \; \mathbb{Z}_N
\nonumber \\
\mu_{k} &\rightarrow& e^{i 2 \pi m k/N} \mu_{k}
\; \; {\rm under} \; \; {\tilde {\mathbb Z}}_N  ,
\label{chargeorderdisorder}
\end{eqnarray}
with $m=0, \ldots, N -1$ and
$\sigma_k$ (respectively $\mu_{k}$) remains unchanged
under the ${\tilde {\mathbb Z}}_N$ (respectively ${\mathbb{Z}}_N$) symmetry.

The ${\mathbb{Z}}_N$ parafermionic CFT is 
generated by the parafermionic
currents $\Psi_{k L}$ and $\Psi_{k R}$ 
($\Psi^{\dagger}_{k L,R} = \Psi_{N-k L,R}$)
which are holomorphic and antiholomorphic fields 
with conformal weights: $h_k = {\bar h}_k = k (N - k)/N$.
They carry respectively a $(k,k)$ and $(k,-k)$ charge under
the ${\mathbb{Z}}_N \times {\tilde {\mathbb{Z}}}_N$ symmetry
and they transform as follows under the KW duality symmetry:
\begin{equation}
\Psi_{k L} \rightarrow \Psi_{k L}, \Psi_{k R} \rightarrow 
\left(-1\right)^k \Psi^{\dagger}_{k R} .
\label{KWparafer}
\end{equation}
In addition, 
these currents satisfy the parafermionic algebra which is defined by the 
following OPEs \cite{para}:
\begin{eqnarray}
\Psi_{k L}\left(z\right) \Psi_{q L}\left(\omega\right)
&\sim& c_{k,q} \left(z-\omega\right)^{-2 kq/N} 
\Psi_{k+q L}\left(\omega\right) \; , \; \; 
{\rm for}  \; \; k+q < 4
\label{opepara1} \\
\Psi_{k L}\left(z\right) \Psi_{q L}^{\dagger}\left(\omega\right)
&\sim& c_{k,N-q} \left(z-\omega\right)^{-2 q(N-k)/N}
\Psi_{k-q L}\left(\omega\right)
\label{opepara2}
\\
\Psi_{k L}\left(z\right) \Psi_{k L}^{\dagger}\left(\omega\right)
&\sim& \left(z-\omega\right)^{-2 k(N-k)/N} \left(1 + \frac{2 h_k}{c}
\left(z-\omega\right)^2 T\left(\omega\right)\right),
\label{opepara3}
\end{eqnarray}
with similar results for the right parafermionic currents.
In Eq. (\ref{opepara3}), $T(z)$ denotes the stress-energy tensor of 
the ${\mathbb{Z}}_N$ CFT and 
the numerical coefficients are given by:
\begin{equation}
c^2_{k,q} = \frac{\Gamma\left(k+q+1\right)
\Gamma\left(5-k\right) \Gamma\left(5-q\right)}{
\Gamma\left(k+1\right) \Gamma\left(q+1\right)
\Gamma\left(6-k-q\right) \Gamma\left(5\right)}.
\label{coeff}
\end{equation}
Apart from these fields, the ${\mathbb{Z}}_N$ CFT also contains 
neutral fields, i.e. invariant under the ${\mathbb{Z}}_N 
\times {\tilde {\mathbb{Z}}}_N$ symmetry, $\epsilon_j$ ($j=1,\ldots, [N/2]$)
with scaling dimension $D_j = 2 j(j+1)/(N+2)$ which are the thermal 
operators of the theory.

It is well known that the ${\mathbb{Z}}_N$ CFT  can be 
viewed as a coset (see for instance Refs. \cite{para,dms}):
$\mathbb{Z}_N \sim$ SU(2)$_N$/U(1)$_N$ or 
$\mathbb{Z}_N \sim$ [SU($N$)$_1$ $\times$ SU($N$)$_1$]/SU($N$)$_2$. 
In fact, there is another coset description 
of the ${\mathbb{Z}}_N$ CFT which is crucial for
the low-energy approach of section II:  
$\mathbb{Z}_N \sim$ SU($2N$)$_1$/Sp($2N$)$_1$.
The corresponding character decomposition has been
found by Altschuler \cite{altschuler} and reads as follows:
\begin{equation}
\chi\left(\Lambda_k \right) = \sum_{j=0}^{N} \; B_j^k \chi \left(\lambda_j\right) 
= \sum_{j=0}^{N} \; \eta \; c^j_k \chi \left(\lambda_j\right),
\label{chardecomp}
\end{equation}
where $\chi\left(\Lambda_k \right), k=0, \ldots, 2N -1$ (respectively 
$\chi \left(\lambda_j\right)$) are
the
characters of the conformal tower corresponding to the SU($2N$)$_1$ 
(respectively Sp($2N$)$_1$) primary fields
$\Phi^{(k)}$ (respectively $\phi^{(j)}$) with scaling dimensions: 
$k(2N-k)/2N$ (respectively $j(2N+2 -j)/2 (N+2)$). 
In Eq. (\ref{chardecomp}), $B_j^k$ are the branching functions 
of the coset which identify to the level-$N$ string functions $c^l_m$
of the SU(2)$_N$ current algebra \cite{gepner,dms}.
The latter are related to the branching functions $b^{j}_{k}$
of the coset $\mathbb{Z}_N \sim$ SU(2)$_N$/U(1)$_N$: 
$b^{j}_{k} = \eta c^{j}_{k}$ ($\eta$ being the Dedekind function)
with the constraint $j \equiv k \; (2)$ \cite{gepner}.

The identification (\ref{chardecomp}) and the 
non-Abelian bosonization rule (\ref{fernonabelboso})
enable one, in principle, to express any fermionic 
operators of the U($2N$)$_1$ CFT
in terms of a U(1) charge boson field and 
operators of the Sp($2N$)$_1 \times {\mathbb{Z}}_N$ CFT. 
Let us first consider the example of the 
representation of the U($2N$)$_1$
field: $L^{\dagger}_{\alpha} R_{\alpha}$. 
Using Eq. (\ref{fernonabelboso}), we have:
\begin{equation}
L^{\dagger}_{\alpha} R_{\alpha} \sim :\exp \left(i\sqrt{2 \pi/N} \Phi_c\right): 
{\rm Tr} \; g.
\label{trgrep}
\end{equation}
The next step of the approach is to use 
the decomposition (\ref{chardecomp}) 
with $k=1$.
The Sp($2N$)$_1$ primary field $\phi^{(1)}$, which transforms
in the fundamental representation (self-conjugate) of Sp($2N$), will 
appear as well as a ${\mathbb{Z}}_N$ field with scaling dimension 
$(N-1)/ N(N+2)$ i.e. $\sigma_1$ or $\mu_1$. 
One way to resolve this ambiguity is to use the parity symmetry $P$ 
in the continuum limit.
Under the parity, $(L,R)_{\alpha} \rightarrow (R,L)_{\alpha}$ and 
$\Phi_{c L,R} \rightarrow - \Phi_{c R,L}$ so that 
${\rm Tr} \; g \rightarrow {\rm Tr} \; g^{\dagger}$ from Eq. (\ref{trgrep}).
However, it is known that, under $P$, 
$\sigma_k$ remains invariant while $\mu_k \rightarrow \mu^{\dagger}_k$ \cite{para}.
We thus deduce the following correspondence:
\begin{equation}
L^{\dagger}_{\alpha} R_{\alpha}
 \sim :\exp \left(i\sqrt{2 \pi/N} \Phi_c\right): {\rm Tr} \; g  \sim 
:\exp \left(i\sqrt{2 \pi/N} \Phi_c\right): \mu_1 {\rm Tr} \; \phi^{(1)} .
\label{grepapp}
\end{equation}
Another interesting operator is the Sp($2N$) singlet 
$L^{\dagger}_{\alpha} {\cal J}_{\alpha \beta} R^{\dagger}_{\beta}$ field
whose representation can be obtained 
from Eq. (\ref{grepapp}) with help of the duality
symmetry (\ref{dualityfer}) of section II.
Since the latter symmetry corresponds to the Gaussian duality 
and the KW transformation of the 
${\mathbb{Z}}_N$ Ising model, i.e. $\sigma_k \leftrightarrow \mu_k$,
we have:
\begin{equation}
L^{\dagger}_{\alpha} {\cal J}_{\alpha \beta} R^{\dagger}_{\beta}
 \sim 
:\exp \left(i\sqrt{2 \pi/N} \Theta_c\right): \sigma_1 {\rm Tr} \; \phi^{(1)} ,
\label{grepbisapp}
\end{equation}
which is consistent with the parity symmetry since 
$L^{\dagger}_{\alpha} {\cal J}_{\alpha \beta} R^{\dagger}_{\beta}$ is 
invariant under $P$.

A more complicate example is the identification of 
the operator $M_0 = \epsilon^{\alpha_1 \ldots \alpha_N \beta_1 \ldots \beta_N}
R^{\dagger}_{\alpha_1} \ldots R^{\dagger}_{\alpha_N}  L^{\dagger}_{\beta_1}  
\ldots L^{\dagger}_{\beta_N}$ in the U(1) $\times$
Sp($2N$)$_1$ $\times$ ${\mathbb{Z}}_N$ basis.
This field is an SU($2N$) singlet and plays a crucial role 
in the discussion of the physical properties of 
the spin-$F$ fermionic model (\ref{hubbardS}) since it is the continuum
representation of the MS instability.
Using the non-Abelian bosonization rules (\ref{fernonabelboso}), we first find that:
\begin{equation}
M_0 \sim :\exp \left(i\sqrt{2 \pi N} \Theta_c\right): {\rm Tr} \; \Phi^{(N)} ,
\label{bosomsapp}
\end{equation}
$\Phi^{(N)}$ being the SU($2N$)$_1$ primary field
which transforms into the self-conjugate representation of 
SU($2N$) given by a Young tableau with $N$ boxes and one 
column. The next step of the approach is to
use Eq. (\ref{chardecomp}) with $k=N$.
Let us assume that $N$ is even, i.e. $N = 2n$, so that 
we find in this case:
\begin{equation}
\chi \left( \Lambda_{2n} \right) = 
\eta \sum_{p=0}^{n} c^{2p}_0 \chi \left( \lambda_{2(n-p)} 
\right),
\label{chardecoms}
\end{equation}
where we have used the symmetry property 
of the level-$N$ string functions:
$c^{l}_m = c^{N-l}_{N-m}$ \cite{gepner}.
It is known that $\eta c^{2p}_0$ with $p=0, \ldots, N/2$ 
describe the conformal towers corresponding to 
the $p^{\mbox{th}}$ thermal operators $\epsilon_p$ 
of the ${\mathbb{Z}}_N$ parafermionic CFT \cite{gepner}.
We thus deduce the following identification 
of the MS instability for even $N$:
\begin{equation}
M_0 \sim :\exp \left(i\sqrt{2 \pi N} \Theta_c\right): 
\sum_{p=0}^{N/2} a_p \; \epsilon_{N/2 -p} \; {\rm Tr} \; \phi^{(2p)} ,
\label{idenmsnevenapp}
\end{equation}
$a_p$ being some numerical coefficients which are not
important in what follows.
When $N$ is odd, a similar approach gives:
\begin{equation}
M_0 \sim :\exp \left(i\sqrt{2 \pi N} \Theta_c\right):
\sum_{p=0}^{(N-1)/2} a_p \; 
\epsilon_{(N-1)/2 -p} \; {\rm Tr} \; \phi^{(2p+1)} .
\label{idenmsnoddapp}
\end{equation}

Finally, we end this Appendix by mentioning 
an interesting identification
of the first parafermionic current 
$\Psi_{1 R,L}$ which generates the 
algebra (\ref{opepara1}, \ref{opepara2}, \ref{opepara3}). 
First, we observe that
$L_{\alpha}^{\dagger} {\cal J}_{\alpha \beta} L_{\beta}^{\dagger}$
and $R_{\alpha}^{\dagger} {\cal J}_{\alpha \beta} R_{\beta}^{\dagger}$
are chiral Sp($2N$) singlets so that only 
the identity operator in the Sp($2N$) sector
can occur in their representation.
We then find the following identifications:
\begin{eqnarray}
L_{\alpha}^{\dagger} {\cal J}_{\alpha \beta} L_{\beta}^{\dagger}
&\simeq& \frac{\sqrt{N}}{\pi}
:\exp \left( i \sqrt{8 \pi/N} \; \Phi_{cL} \right): \Psi_{1 L}
\nonumber \\
R_{\alpha}^{\dagger} {\cal J}_{\alpha \beta} R_{\beta}^{\dagger}
&\simeq& \frac{\sqrt{N}}{\pi}
:\exp \left( -i \sqrt{8 \pi/N} \; \Phi_{cR} \right): \Psi_{1 R} ,
\label{pararepapp}
\end{eqnarray}
which satisfy the parafermionic algebra (\ref{opepara3}) for $k=1$.
The representation of the parafermionic 
current $\Psi_{k R,L}$ with $k \ge 2$ can then also
been derived from the correspondence (\ref{pararepapp}) and 
the application of the parafermionic algebra (\ref{opepara1},\ref{opepara2}).



\begin{thebibliography}{101}
\bibitem{hofstetter} 
W. Hofstetter, J. I. Cirac, P. Zoller, E. Demler, M. D. Lukin,
Phys. Rev. Lett. 89 (2002) 220407.
\bibitem{reviewcold} For recent reviews, see for instance:
M. Lewenstein, A. Sanpera, V. Ahufinger, B. Damski, A. Sen De,
U. Sen, Advances in Physics 56 (2007) 243;        
I. Bloch, J. Dalibard, W. Zwerger,
arXiv: cond-mat/0704.3011.
\bibitem{jaksch}
D. Jaksch, C. Bruder, J. I. Cirac, C. W. Gardiner, P. Zoller,
Phys. Rev. Lett. 81 (1998) 3108. 
\bibitem{greiner}
M. Greiner, O. Mandel, T. Esslinger, T. W. H{\"a}nsch, I. Bloch, 
Nature 415 (2000) 39.
\bibitem{ho}
T.-L. Ho, Phys. Rev. Lett. 81 (1998) 742.
\bibitem{ohmi}
T. Ohmi, K. Machida, J. Phys. Soc. Jpn. 67 (1998) 1822.
\bibitem{zhou}
E. Demler, F. Zhou, Phys. Rev. Lett. 88 (2002) 163001;
F. Zhou, G. W. Semenoff, {\it ibid.} 97 (2006) 180411;
G. W. Semenoff, F. Zhou, {\it ibid.} 98 (2007) 100401;
J. L. Song, G. W. Semenoff, F. Zhou, {\it ibid.} 98 (2007)
160408.
\bibitem{moore}
S. Mukerjee, C. Xu, J. E. Moore, Phys. Rev. Lett. 97 (2006) 120406.
\bibitem{demler}
R. Barnett, A. Turner, E. Demler, Phys. Rev. Lett. 97 (2006) 180412;
Phys. Rev. A  76 (2007) 013605;
A. Turner, R. Barnett, E. Demler,  A. Vishwanath, Phys. Rev. Lett. 98 (2007) 190404.
\bibitem{yip}
S. K. Yip, Phys. Rev. A 75 (2007) 023625.
\bibitem{bosonspin3}
L. Santos, T. Pfau, Phys. Rev. Lett. 96 (2006) 190404;
R. B. Diener, T.-L. Ho, {\it ibid.} 96 (2006) 190405.
\bibitem{bosonspin1}
D. M. Stamper-Kurn, M. R. Andrews, A. P. Chikkatur, 
S. Inouye, H.-J. Miesner, J. Stenger, W. Ketterle,
Phys. Rev. Lett. 80 (1998) 2027;
J. Stenger,  S. Inouye, D. M. Stamper-Kurn, 
H.-J. Miesner, A. P. Chikkatur,
W. Ketterle, Nature (London) 396 (1998) 345;
M. S. Chang {\it et al.}, Nature Phys. 1 (2005) 111.
\bibitem{spin3}
A. Griesmaier, J. Werner, S. Hensler, J. Stuhler, T. Pfau,
Phys. Rev. Lett. 94 (2005) 160401.
\bibitem{hoyip}
T.-L. Ho, S. Yip, Phys. Rev. Lett. 82 (1999) 247.
\bibitem{honerkamp}
C. Honerkamp, W. Hofstetter, Phys. Rev. Lett. 92 (2004) 170403;
Phys. Rev. B 70 (2004) 094521.
\bibitem{paananen}
T. Paananen, J. P. Martikainen, P. T{\"o}rm{\"a}, 
Phys. Rev. A 73 (2006) 053606;
T. Paananen, P. T{\"o}rm{\"a}, J. P. Martikainen,
{\it ibid.} 75 (2007) 023622.
\bibitem{cherng}
R. W. Cherng, G. Refael, E. Demler, Phys. Rev. Lett. 99 (2007) 130406.
\bibitem{he}
L. He, M. Jin, P. Zhuang, Phys. Rev. A 74 (2006) 033604.
\bibitem{schuck}
G. R\"opke, A. Schnell, P. Schuck, P. Nozi\`eres,
Phys. Rev. Lett.  80 (1998) 3177;
G. R\"opke, P. Schuck,  Mod. Phys. Lett. A 21 (2006) 2513.
\bibitem{tohsaki}
A. Tohsaki, H. Horiuchi, P. Schuck, G. R\"opke,
Phys. Rev. Lett. 87 (2001) 192501.
\bibitem{deanlee}
D. Lee,  Phys. Rev. Lett. 98 (2007) 182501.
\bibitem{nozieres}
P. Nozi\`eres, D. Saint James, J. Phys. (Paris) 43
(1982) 1133.
\bibitem{doucot}
B. Dou\c{c}ot, J. Vidal,
Phys. Rev. Lett. 88 (2002) 227005.
\bibitem{schneider}
C. W. Schneider, G. Hammerl, G. Logvenov, T. Kopp,
J. R. Kirtley, P. J. Hirschfeld, J. Mannhart, 
Europhys. Lett. 68 (2004) 86.
\bibitem{aligia}
A. A. Aligia, A. P. Kampf, J. Mannhart,
Phys. Rev. Lett. 94 (2005) 247004; 
L. M. L. Hilario, A. A. Aligia, arXiv: cond-mat/0702143.
\bibitem{affleckbipairing}
M. S. Chang, I. Affleck, Phys. Rev. B 76 (2007) 054521. 
\bibitem{miyake}
H. Kamei, K. Miyake,
J. Phys. Soc. Jpn. 74 (2005) 1911;
A. S. Stepanenko, J. M. F. Gunn,
arXiv: cond-mat/9901317.
\bibitem{Wu2005}
C. J. Wu, Phys. Rev. Lett. 95 (2005) 266404.
\bibitem{phle}
P. Lecheminant, E. Boulat, P. Azaria,
Phys. Rev. Lett. 95 (2005) 240402.
\bibitem{capponiMS}
S. Capponi, G. Roux, P. Lecheminant, P. Azaria, E. Boulat, 
S. R. White, Phys. Rev. A 77 (2008) 013624.
\bibitem{Rapp}
A. Rapp, G. Zar\'and,
C. Honerkamp, W. Hofstetter, Phys. Rev. Lett. 98 (2007) 160405;
A. Rapp, W. Hofstetter, G. Zar\'and, arXiv: cond-mat/0707.237.
\bibitem{schlottmann}
X. W. Guan, M. T. Batchelor, C. Lee, 
H. Q. Zhou, arXiv: cond-mat/0709.1763;
X. Liu, H. Hu, P. D. Drummond, arXiv: cond-mat/0709.2273;
P. Schlottmann, J. Phys. Condens. Matter 6
(1994) 1359.
\bibitem{sachdev} N. Read, S. Sachdev, Phys. Rev. Lett. 66
(1991) 1773.
\bibitem{wusp}
C. Wu, S.-C. Zhang, Phys. Rev. B 71
(2005) 155115.
\bibitem{zhang}
C. J. Wu, J. P. Hu,
S.-C. Zhang, Phys. Rev. Lett. 91 (2003) 186;
C. J. Wu, Mod. Phys. Lett. B 20 (2006) 1707.
\bibitem{para} A. B. Zamolodchikov, V. A.  Fateev, Sov. Phys. JETP
62 (1985) 215.
\bibitem{tsvelikbook} A. M. Tsvelik,
\textit{Quantum Field Theory in Condensed Matter Physics}
(Cambridge University Press, Cambridge, 1995).
\bibitem{bookboso}
A. O.~Gogolin, A. A.~Nersesyan, A. M.~Tsvelik,
\textit{Bosonization and Strongly Correlated Systems}
(Cambridge university press, UK, 1998).
\bibitem{giamarchi}
T. Giamarchi, \textit{Quantum Physics in One Dimension}
(Clarendon press, Oxford, UK, 2004).
\bibitem{witten}
E. Witten, Commun. Math. Phys. 92 (1984) 455.
\bibitem{knizhnik}
V. G. Knizhnik and A. B. Zamolodchikov, Nucl. Phys. B 247 (1984) 83. 
\bibitem{affleckspinchain} I. Affleck, Nucl. Phys. B 265 (1986) 409;
I. Affleck, F. D. M. Haldane, Phys. Rev. B 36 (1987) 5291.
\bibitem{dms}
P. Di Francesco, P. Mathieu, D. S\'en\'echal,
\textit{Conformal Field Theory} (Springer, Berlin, 1997).
\bibitem{aza4spin}
P. Azaria, A. O. Gogolin, P. Lecheminant, A. A. Nersesyan,
Phys. Rev. Lett. 83 (1999) 624;
P. Azaria, E. Boulat, P. Lecheminant,
Phys. Rev. B 61 (2000) 12112;
P. Lecheminant, K. Totsuka, {\it ibid.}  71  (2005) 020407(R);
{\it ibid.}  74  (2006) 224426.
\bibitem{controzzi}
D. Controzzi, A. M. Tsvelik,
Phys. Rev. Lett. 95 (2006) 097205;
Phys. Rev. B  72 (2005) 035110.
\bibitem{Affleck-G-S-Z-89}
I. Affleck, D. Gepner, H. J. Schulz, T. Ziman,
J. Phys. A: Math. Gen. 22 (1989) 511;
R. R. P. Singh, M. E. Fisher,  R. Shankar,
Phys. Rev. B 39 (1989) 2562.
\bibitem{majumdar}
K. Majumdar, M. Mukherjee,
J. Phys. A: Math. Gen.  35 (2002) L543.
\bibitem{affleck}
I.~Affleck, Nucl. Phys. B 305 (1988) 582.
\bibitem{assaraf}
R. Assaraf, P. Azaria, M. Caffarel,  P. Lecheminant,
Phys. Rev. B  60 (1999) 2299.
\bibitem{gross}
D. J. Gross, A. Neveu, Phys. Rev. D 10 (1974) 3235.
\bibitem{andrei}
N.~Andrei, J. H.~Lowenstein, Phys. Lett. B  90 (1980) 106.
\bibitem{konik} R. Konik, H. Saleur, A. W. W. Ludwig,
Phys. Rev. B 66 (2002) 075105.
\bibitem{balents}
H.-H.~Lin, L.~Balents, M. P. A.~Fisher,
Phys. Rev. B  58 (1998) 1794.
\bibitem{boulat} R. Assaraf, P. Azaria, E. Boulat, M. Caffarel,
P. Lecheminant, Phys. Rev. Lett.  93 (2004) 016407.
\bibitem{affleckondo}
I. Affleck, Acta Polonica B 26, 1869 (1995).
\bibitem{altschuler} D. Altschuler, Nucl. Phys. B 313 (1989) 293.
\bibitem{seidel} 
A. Seidel, D. H. Lee,
Phys. Rev. Lett. 93 (2004) 046401;
Phys. Rev. B 71 (2005) 045113. 
\bibitem{capponi}
S. Capponi, G. Roux, P. Azaria, E. Boulat, P. Lecheminant,
Phys. Rev. B 75 (2007) 100503(R).
\bibitem{fateev} V. A. Fateev, Int. J. Mod. Phys. A 6  2109
(1991).
\bibitem{sutherland} 
B. Sutherland, Phys. Rev. B 12 (1975) 3795.
\bibitem{orignac}
E. Orignac, R. Citro,  Eur. Phys. J. B 33 (2003) 419.
\bibitem{capponipreprint}
S. Capponi {\it et al.}, in preparation.
\bibitem{leclair}
B. Gerganov, A. LeClair, M. Moriconi,
Phys. Rev. Lett. 86 (2001) 4753; 
A. LeClair, Phys. Rev. B 64 (2001) 045329. 
\bibitem{gepner}
D. Gepner, Z Qiu, Nucl. Phys. B 285 (1987) 423.
\end{thebibliography}
\end{document}